\documentclass[twocolumn,showpacs,preprintnumbers,amsmath,amssymb]{revtex4}
\usepackage{graphicx}
\usepackage{dcolumn}
\usepackage{bm}
\begin{document}

\title{The influence of structural disorder on magnetic domain formation in perpendicular anisotropy thin films}

\author{M.S. Pierce$^{1-3}$,  J.E. Davies$^{4,5}$ J.J. Turner$^{6,7}$,  K. Chesnel$^{8}$, E. E. Fullerton$^{9,10}$, J. Nam$^{1}$, R. Hailstone$^{11}$, S.D. Kevan$^{6}$,  J.B. Kortright$^{8}$,  Kai Liu$^{4}$, L.B. Sorensen$^{3}$, B.R. York$^{12}$, and O. Hellwig$^{13}$}

\address{$^{1}$School of Physics and Astronomy, Rochester Institute of Technology, Rochester NY 14623} 

\address{$^{2}$Materials Science Division, Argonne National Laboratory, Argonne Il 60439}

\address{$^{3}$Department of Physics, University of Washington,
Seattle WA 98195}

\address{$^{4}$Department of Physics, University of California, Davis, CA 95616}

\address{$^{5}$Advanced Technology Group, NVE Corporation, Eden Prairie, MN 55344}

\address{$^{6}$Department of Physics, University of Oregon, Eugene
Oregon 97403}

\address{$^{7}$SLAC National Accelerator Laboratory, Menlo Park CA 94025}

\address{$^{8}$Lawrence Berkeley National Laboratory, Berkeley
CA 94720}

\address{$^{9}$Department of Electrical and Computational Engineering, University of California, San Diego, La Jolla CA 92093}
\address{$^{10}$Center for Magnetic Recording Research, University of California, San Diego,  La Jolla CA 92093}
\address{$^{11}$Center for Imaging Science, Rochester Institute of Technology, Rochester NY 14623} 

\address{$^{12}$Materials Laboratory, HGST, a Western Digital Company,
San Jose CA 95193}
\address{$^{13}$San Jose Research Center, HGST, a Western Digital Company,
San Jose CA 95135}

\date{\today}

\begin{abstract}

Using a combination of resonant soft x-ray scattering, magnetometry, x-ray reflectivity and microscopy techniques we have investigated the magnetic properties and microstructure of a series of perpendicular anisotropy Co/Pt multilayer films with respect to structural disorder tuned by varying the sputtering deposition pressure. The observed magnetic changes in domain size, shape and correlation length originate from structural and chemical variations in the samples, such as chemical segregation and grain formation as well as roughness at the surface and interfaces, which are all impacted by the deposition pressure.  All samples exhibited short range ``liquid-like" positional ordering over significant portions of their major hysteresis loops, while only the lowest disorder samples showed evidence of a random ``gas-like" distribution of magnetic domains, present just after nucleation and as well as prior to saturation. The structural and chemical disorder induced by the higher deposition pressure first leads to an increase in the number of magnetic point defects that limit free domain wall propagation. Then, as the sputtering pressure is further increased, the domain wall energy density is lowered due to the formation of local regions with reduced magnetic moment, and finally magnetically void regions appear that confine the magnetic domains and clusters, similar to segregated granular magnetic recording media.


\end{abstract}

\pacs{ 07.85.+n, 61.10.-i, 78.70.Dm, 78.70.Cr}

\maketitle

\section{Introduction
}

Perpendicular anisotropy magnetic thin films can exhibit complex domain structures due to the competition between short range and long range interactions.  Magnetic domain patterns include highly ordered labyrinths, bubbles, fractal patterns, and many more variations.   The different shapes and geometries of the domains are greatly influenced by the degree of lateral magnetic heterogeneity in the system caused by chemical and structural variations present within the thin film itself.  In addition to ferromagnetic systems, competing domains are also found in many other areas of science. Labyrinth, stripe, and bubble domains \cite{seul_science_95} have been observed in a large variety of systems and at many different length scales.  These include synthetic antiferromagnets \cite{olav_jmmm_07}, ferrimagnetic multilayers \cite{olav_apl_2011} and garnet films \cite{seul_pra_92}, nanoscale self-assembled systems \cite{plass_jpcm_02}, ferroelectrics, smectic and lamellar phases of liquid crystals, ferrofluids, block co-polymers and more.  In the case of magnetic systems, the disorder can often be directly attributed to structural and chemical disorder through pinning sites, static frozen random fields, magnetic voids and variation in the lateral magnetic exchange coupling.   

While such magnetic domain patterns are interesting in their own right, understanding the character of microscopic magnetic domains is of vital importance to magnetic data storage technology.  Co/Pt, and similar multilayer films represent some of the best and most frequently studied model systems  with perpendicular anisotropy \cite{davies_jap_11,hannon_prl_88,osgood_jmmm_99,jeff_prb_2001,dudzik_prb_00,zeper_jap_91, donzelli}.  Yet at times surprising gaps in our understanding are uncovered regarding widely accepted magnetic phenomena \cite{berger_prb_10,leighton_phys_10,davies_apl_09}.

Here we present an investigation of the effects of structural and chemical disorder on the lateral inter-grain exchange, and therefore on the magnetic domain structure, in a series of Co/Pt multilayer samples with perpendicular magnetic anisotropy.  The disorder is brought about by increasing the sputter growth pressure \cite{7} of the Co/Pt multilayers  during deposition and has direct effects on the magnetic reversal behavior and microscopic magnetic domain formation. The disorder itself is manifest in an increased surface and interface roughness as well as an increased degree of lateral chemical and structural heterogeneity within the magnetic thin film. This lateral heterogeneity can evolve from simple impurities and point defects that act as nucleation or pinning sites to the formation of a distinct reduced or non-magnetic grain boundary phase in between the magnetic grains.

This investigation draws upon several complementary experimental techniques:  resonant soft x-ray scattering, x-ray reflectivity, and magnetometry.  Additional insight into the interplay of chemical structure and magnetic domains was obtained by atomic force microscopy (AFM), magnetic force microscopy (MFM), and Transmission Electron Micorscopy (TEM).  Resonant scattering from the transition metal L edge, which couples the electron states and photon polarization, is capable of directly probing the magnetic character of thin films \cite{hannon_prl_88,osgood_jmmm_99,jeff_prb_2001, jefftwo, hu_srn_01} and related samples \cite{chesnel_prb_02,beutier_prb_05,ogrin_jap_08}.  Variations in the application of this mechanism continues to produce powerful new magnetic characterization techniques \cite{turner_prl_11, tripathi_pnas_11, konings_prl_11}.  

In our studies, we found all samples to exhibit magnetic domain patterns with short range magnetic ordering along significant portions of their respective major hysteresis loops.  However, our low disorder (low deposition pressure) samples showed dilute phase regions shortly after nucleation, and just prior to saturation, with little or no short range ordering. From our scattering images we have been able to extract the ensemble averaged domain widths and correlation lengths, and study how they depend on the externally applied field.  By combining our x-ray scattering results with magnetometry data we were able to extract the average individual up and down domain widths as they depend upon the applied magnetic field. Additionally, TEM was used to examine the two extremes of our samples (highest and lowest deposition pressure), clearly showing the presence of physical grain boundaries in the high disorder (high deposition pressure) sample and the absence of those features in the low disorder sample.  We show below how the growth pressure affects the amount and kind of disorder in the films and subsequently how this structural disorder impacts the magnetic properties of the samples.

\section{Samples and Structural Characterization.}
Symmetry considerations at interfaces have long been understood as a mechanism to drive perpendicular magnetic anisotropy \cite{neel}. The magnetic properties of multilayer thin films \cite{denbroeder_prl_88}, among them Co/Pt \cite{zgli_jap_92, bain_jap_93}, are known to depend upon microstructure, chemical segregation, and other effects introduced at the time of sample deposition. Previously we have explored the magnetic memory effects of these samples and many of these experimental details are already published \cite{pierce_prl_05, pierce_prb_07}.  In this section we revisit and summarize the essential findings of the previous studies, but incorporate new, additional structural characterization results not previously published.

\begin{figure*}
\includegraphics[width=17cm]{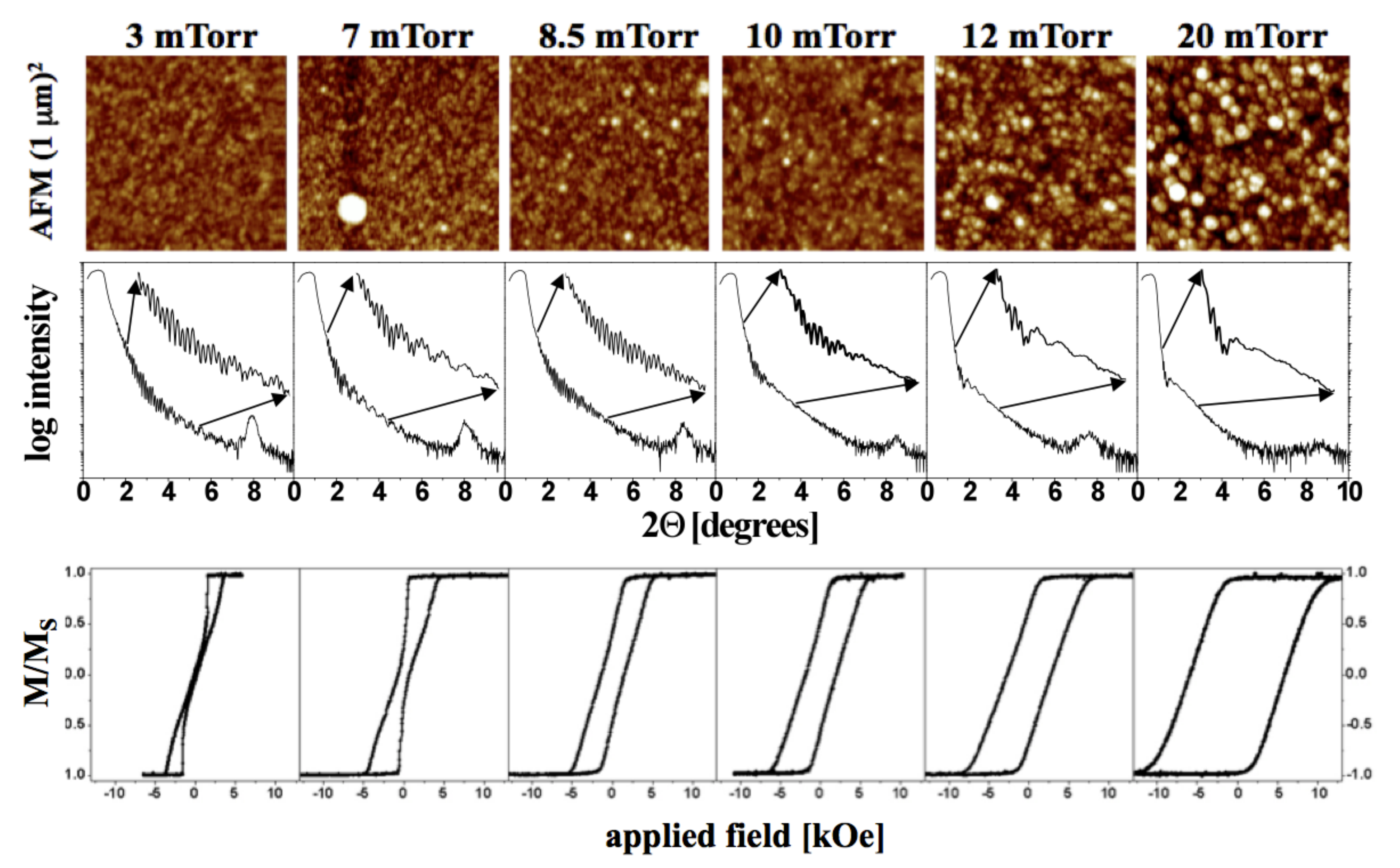}
\caption{(color online) AFM, x-ray reflectivity, and magnetometry data measured for the 6 pressure series samples.  The AFM images show the sample surface becoming increasingly rough due to the formation of distinct grains as the sputtering pressure increases.   In order to have a good comparison among different pressure samples all AFM images were taken with the same AFM tip and have the same vertical height color scale. The reflectivity curves, obtained using Cu K$_{\alpha}$ radiation, reveal an increase in surface and interface roughness, as well as the disappearance of the multilayer super structure peak with increased deposition pressure. Both the AFM and reflectivity data were used to determine the overall surface roughness of the samples. The MOKE hysteresis loops, measured in perpendicular geometry, show a continuous increase in coercivity with a change in hysteresis loop shape occurring between 7 and 8.5 mTorr.}
\label{fg:hysteresis_loops} 
\end{figure*}

Our thin-film samples were grown on smooth, low-stress, 160-nm-thick SiN membranes by
magnetron sputtering. The samples had 20-nm-thick Pt buffer layers,
and 2.3-nm-thick Pt caps preventing oxidation. Between the buffer layer and the cap, the samples had 50 repeating units of a 0.4-nm-thick Co layer and a 0.7-nm-thick Pt layer. While the six samples had nominally identical multilayer structure, they were grown at different argon sputtering pressures to vary systematically the degree of disorder in the samples. During growth, we adjusted the deposition time to keep the Co and Pt layer thickness constant over the entire series  \cite{7}.  The samples were grown at six different pressures: 3, 7, 8.5, 10, 12, and 20 mTorr.  We identify each sample based on its growth pressure.  AFM, X-ray reflectivity and magnetometry results for the complete pressure series as presented in earlier publications \cite{pierce_prl_05, pierce_prb_07} are summarized in Fig.\ref{fg:hysteresis_loops}. We will re-discuss these earlier findings in the following sections, while incorporating new and additional experimental data and results, in order to complete our understanding of structural disorder formation originating from deposition pressure variation on the one hand and its impact on magnetic domain formation on the other hand.

\subsection{X-ray Reflectivity and Diffraction}

\begin{figure}
\includegraphics[width=8.65cm]{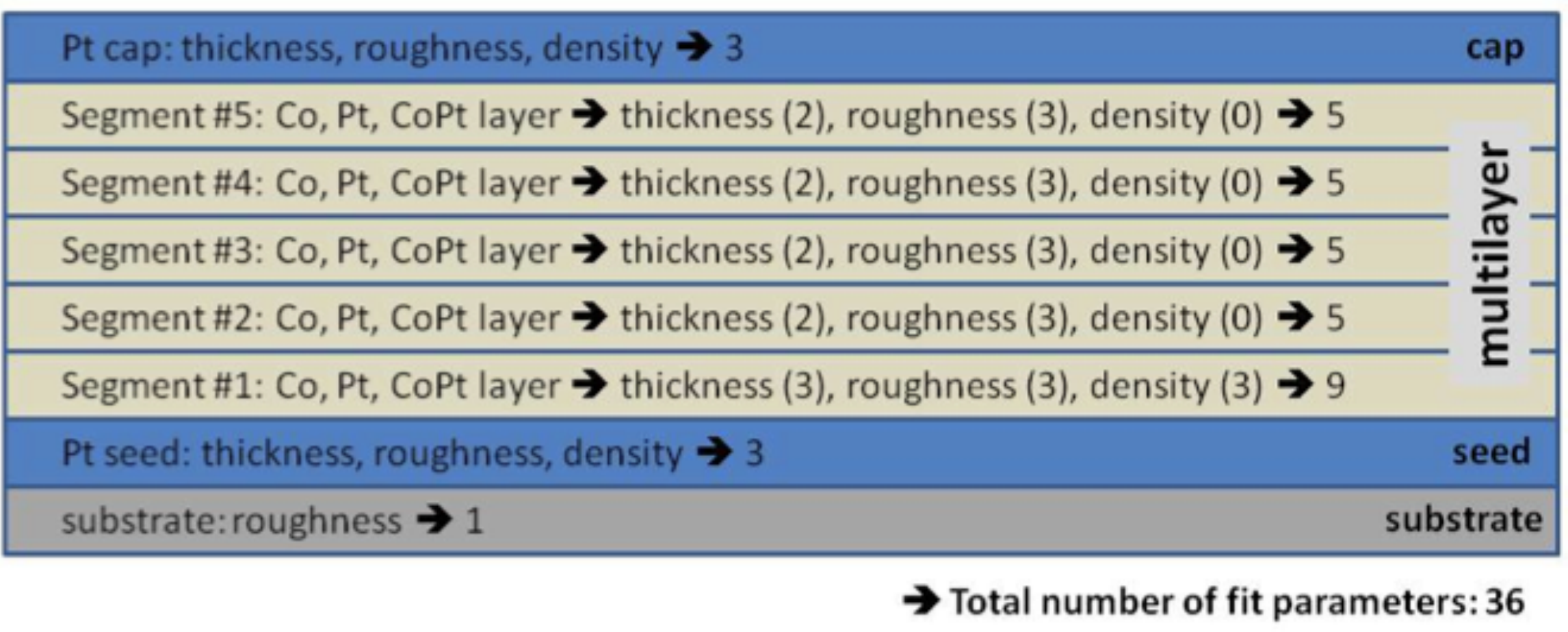}
\includegraphics[width=8.5cm]{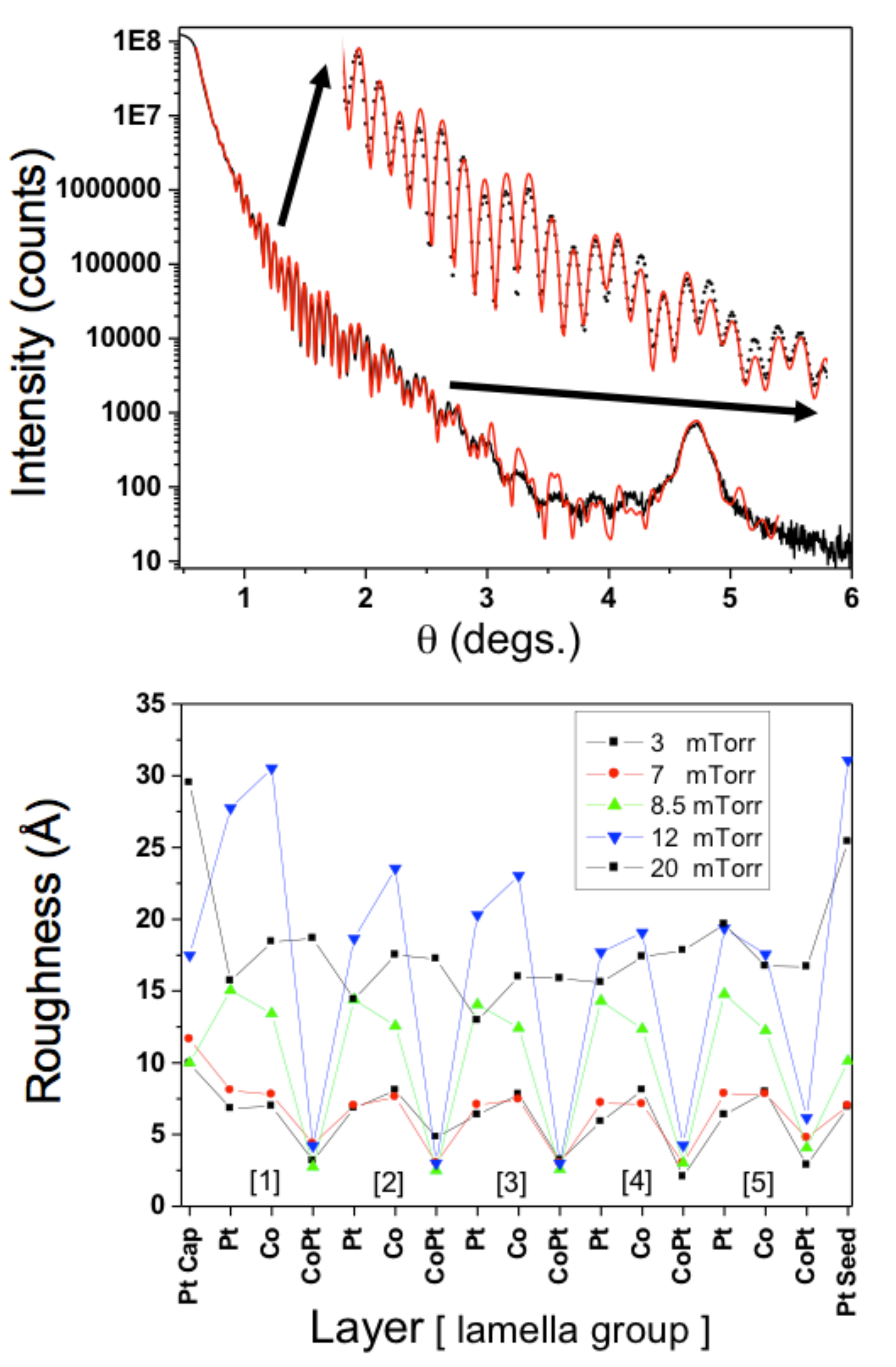}
\caption{(Color Online) Shown at the top is an illustration of the model that was used for fitting the XRR data with a total number of 36 free parameters. In the middle panel we display a fit to the 3 mTorr reflectivity data as obtained from the model shown above.  The bottom panel shows the corresponding roughness profiles extracted from the reflectivity data for different deposition pressure samples.}
\label{fg:reflectivity}
\end{figure}

After the initial x-ray reflectivity (XRR) measurements with Cu K$_{\alpha}$ radiation, as shown in Fig.1, we decided to collect higher resolution reflectivity data using Co K$_{\alpha}$ radiation for fitting the pressure dependent changes in our samples. The higher resolution XRR data (an example is shown in Fig. 2 in the middle graph) for each of the films was then modeled in 5 groups of layers or lamella segments, sandwiched between the Pt seed and Pt capping layer using the BEDE REFS 4 analysis package \cite{schug_xray}. The general XRR fitting model used here is illustrated and outlined in the top panel of Fig. \ref{fg:reflectivity} . Instead of treating every multilayer repeat individually, we introduced lamella segments in order to significantly reduce the number of parameters required for a reasonable fit.  As shown in the figure, this model still requires 36 free parameters.  Each lamella segment contains 10 repeat layers of Co/Pt, where the Co and Pt layer thickness and roughness were held fixed within each segment. Specular XRR cannot easily separate roughness from intermixing at the interfaces. However, simple bilayer Co/Pt laminate models, showed much lower Pt, and higher Co densities in the laminate by nearly 20\% from that found in the Pt seed and capping layers or thicker Co layers, indicating significant Co/Pt intermixing. As such, an interfacial layer was added for each Co/Pt repeated layer in an attempt to describe Co atom intermixing into existing Pt surfaces within each lamella segment in the model presented here. The density of the Co, Pt, and interfacial CoPt layer, as well as the thickness of the CoPt interlayer, were each fit to one respective value that in our model was used for the complete multilayer. However the Co and Pt thickness as well as the Co, Pt and CoPt interlayer roughness were held fixed only within each segment, but were allowed to vary across the 5 different lamella segments that form the complete multilayer structure in order to describe the thickness and roughness evolution with increasing stack thickness or deposition time.  It was necessary to include this type of lamella segment - XRR modeling in order to describe the experimentally observed changes within the reflectivity profiles i.e. the broadening and disappearance of both the repeat layer super-lattice peak around 8.1 degrees and the interference fringes as evolving with increasing deposition pressure. While we observe some variation in the Co and Pt layer thicknesses that lead to multilayer repeat variations from 0.93 to 1.2 nm  for the different lamella segments, we do not obtain a systematic trend in Co, Pt or multilayer period thickness with increasing deposition time as we form the multilayer stack. 

In the lowest disorder 3 mTorr sample, the super-lattice peak is clearly visible, corresponding to a well defined layer thickness of 1.09 nm $\pm$ .05 nm.  For all pressures explored here the superlattice peak remains at the same position within these error bars, which is also consistent with all the average multilayer periods obtained from the fits within our model. For the 3 mTorr sample the superlattice peak has already broadened from a perfect multilayer by small thickness and roughness changes among the collective lamella segments \cite{7}. However, in the more disordered samples the super-lattice peak broadens even further, indicating higher interfacial roughness and/or intermixing of the Co and Pt laminate layers \cite{takano_jap_00}. The intermediate frequency fringes present in the low disorder samples are from the Pt seed, and the high frequency fringes are from the combined thickness of the Pt cap, Pt seed and Co/Pt multilayer.

Using the lamella segment model described above and illustrated in the top of Fig. \ref{fg:reflectivity}, it was possible to refine the reflectivity data and extract the roughness depth profiles for the various deposition pressures.  Shown in Fig. 2 is an example of the fit of the 3 mTorr sample reflectivity (middle) and the comparison to the extracted interfacial roughness for all the samples (bottom). For the two lowest depositions pressure samples, the roughness is found to be relatively small throughout the multilayers. The 8.5 and 12 mTorr \cite{by_ref_note} samples show an increase in roughness through out the multilayer structure, but not yet evidence of strong CoPt inter-alloying, thus maintaining the characteristics of the multilayering.   From large angle x-ray diffraction analysis we find that the Co/Pt laminate and Pt seed/capping layer strain increases with deposition pressure, until a point is reached at 20 mTorr in which the laminate cannot maintain a (111) growth  texture anymore and it is energetically favorable to grow in a more randomly oriented relaxed polycrystalline structure.  At this point the XRR modeling shows a more conformal roughness of this albeit very thin interfacial layer accompanied by an almost complete loss of the superlattice peak intensity.  This results in a relaxation of the Pt seed/cap strain approaching again the same strain state as originally observed at low deposition pressures, whereas the Pt laminate still retains some pressure induced growth strain.  The multilayer structure is almost not recognizable any more even though the system maintains overall perpendicular anisotropy.  It is worth noting that the behavior of the surface roughness parameter extracted from the reflectivity is consistent with the independent atomic force microscopy (AFM) measurements across the pressure series, thus confirming independently the validity of our reflectivity lamella segment model.  Both techniques, AFM and XRR, yield a surface roughness evolution that increases by a factor of 3 from the lowest deposition pressure 3 mTorr sample to the highest deposition pressure 20 mTorr sample. All AFM values are listed in Table 1.  While the relative roughness trends of AFM and XRR analysis across the pressure series match very well, absolute XRR roughness values are about a factor of 2 larger than corresponding AFM values. This may be due to the fact that AFM measures physical surface roughness only, while XRR also measures density variations towards the surface due to intermixing, grain formation or relaxation. In that sense the XRR roughness is always expected to be larger than the corresponding AFM roughness.

In addition to the small angle reflectivity measurements we also performed large angle x-ray experiments to monitor the crystalline microstructure changes via analysis of the Bragg reflections for the various deposition pressures. Corresponding results collected with Cu K$_{\alpha}$ radiation are presented in Figure 3, where we display the Bragg reflections as obtained from the Pt seed layer and the Co/Pt multilayer. Both layers are dominated by a Pt face-centered cubic (FCC) structure with their closest packed (111) direction oriented out-of-plane. As expected the Bragg position of the Pt seed layer matches very closely the Pt bulk lattice parameter, while the Bragg position of the Co/Pt multilayer is shifted to larger angles i.e. smaller lattice spacing due to the Co interlayers within the multilayer. With increasing deposition pressure both Bragg peaks shift towards larger angles, thus indicating the build up of tensile lattice strain in the out-of-plane direction. However when increasing the deposition pressure beyond 12 mTorr towards 20 mTorr the stress built up within the multilayer is suddenly relaxed (multilayer Bragg peak shifts back to smaller angles) as the (111) texture is lost completely due to the reduced kinetic energy available of the arriving atoms at high deposition pressure \cite{windischmann}. This is confirmed by comparing rocking curve profiles of the 3 and 20 mTorr samples in the inset of Figure 3 that clearly show a transformation of a well textured FCC (111) multilayer at 3 mTorr into a polycrystalline layer structure at 20 mTorr deposition pressure.

\begin{figure}
\includegraphics[width=8.5cm]{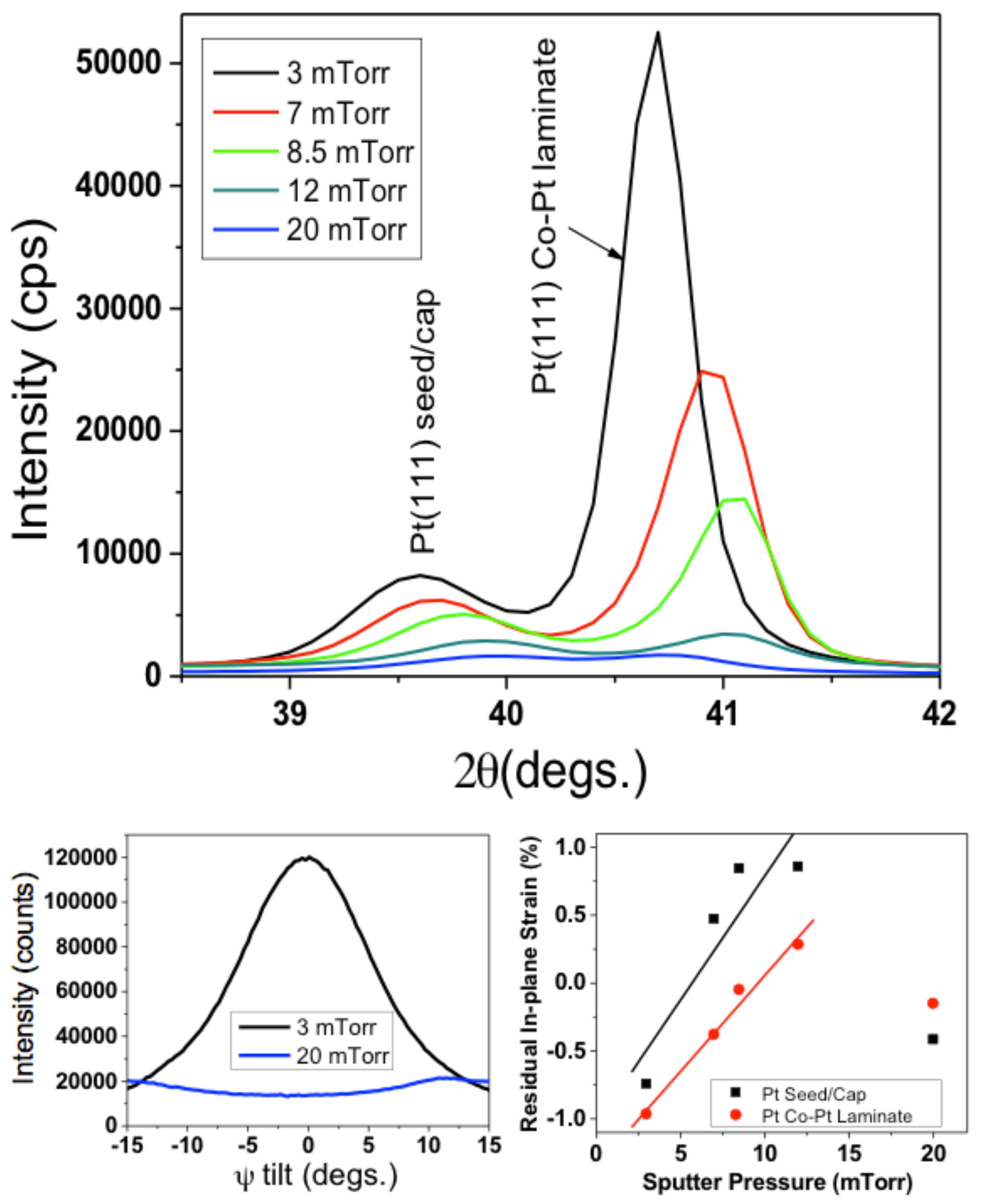}
\caption{(Color Online) Large-angle Bragg profiles for the various deposition pressures are shown in the top panel. The Pt underlayer and the Co/Pt multilayer reveal a strained FCC (111) out-of-plane orientation. The lower left panel compares Co/Pt multilayer rocking scan profiles of the 3 and 20 mTorr sample and confirms the gradual loss of FCC (111) out-of-plane texture towards a completely polycrystalline structure as the deposition pressure is increased. The intensity increase for the 20 mTorr rocking curve at $\pm$15 degrees is due to Bragg reflection tails that originate from the Si substrate. The lower right panel shows the residual in-plane strain of the Pt/Co laminate and Pt Seed/Cap layers for the samples as a function of deposition pressure.  The strain increases in the samples from 3 to 12 mTorr and then relaxes for the 20 mTorr sample due to a loss of the (111) texture.}
\label{fg:xrd}
\end{figure}

\subsection{Transmission Electron Microscopy}

TEM is a well established technique for examining nano-scale objects including magnetic structures \cite{tanahasi_ieee_09, doerner_ieee_01, toney_jap_03} that is quite complementary to the x-ray characterization presented in the previous section.  
For our studies we used the Rochester Institute of Technology NanoImaging Laboratory TEM facilities. Prior to imaging the samples were etched by floating on KOH solution to remove the substrate and base layer.  Etching times were varied from 15 minutes to 3 hours to ensure that the results would be independent of the etching time.

\begin{figure}
\includegraphics[width=9cm]{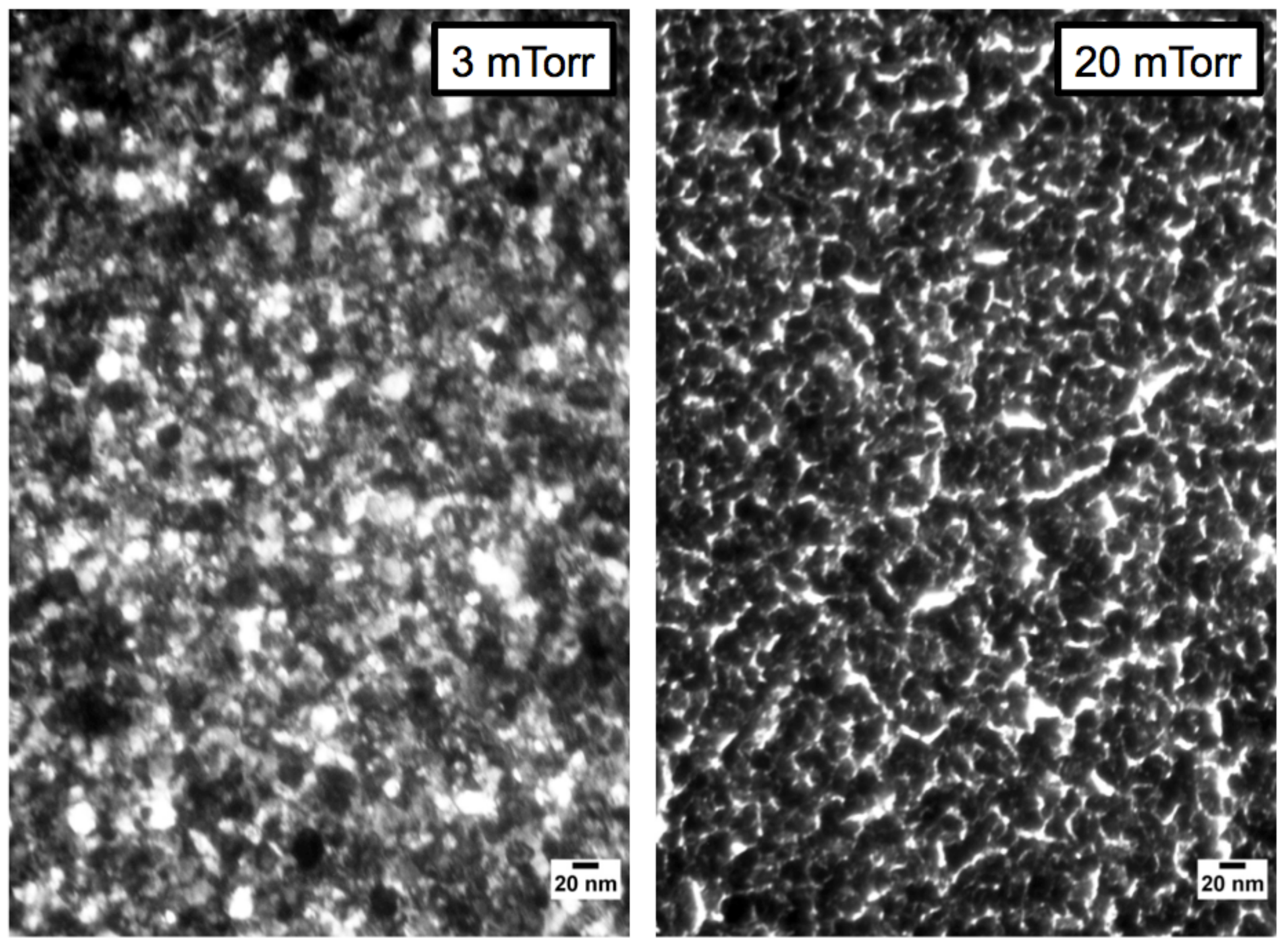}
\caption{TEM images for the 3 mTorr (left) and 20 mTorr (right) samples.  The 3 mTorr sample shows structural grains randomly distributed and in direct contact with each other, without any distinct grain boundary phase present.  The 20 mTorr sample exhibits clear formation of a distinct grain boundary phase, which contains less dense material than the surrounding grains.
}
\label{fg:tem}
\end{figure}

Representative plane-view TEM images are shown in Fig. \ref{fg:tem} and clearly demonstrate the existence of distinct grain boundary phase formation in the 20 mTorr sample that are not present in the 3 mTorr sample. The bright areas within the 20 mTorr sample are regions of relatively high electron transparency in between the grains with an average width of $\sim 10$ nm. Elemental analysis of the 20 mTorr sample using energy-dispersive x-ray microanalysis did not reveal any significant chemical compositional phase separation at the boundaries.  Instead the boundaries, while substantially less dense or thinner than the surrounding grains, contained the same relative composition ratio of Co and Pt. The 3 mTorr sample does show significant grain structure as well, however the grains are here uniformly distributed and do not have a specific lower density boundary phase in-between them. It is possible that etching, while appearing relatively time independent, does effect the samples differently.  While the images were collected with the same instrumental contrast, differences in the apparent contrast could be due to differences in thickness at the time of TEM measurement.
 
The presence of the elongated low density structures in the high pressure 20 mTorr sample, and the corresponding absence of such structures in the 3 mTorr samples, confirm the formation of a grain boundary phase with reduced Co/Pt multilayer density and thus reduced or even vanishing magnetization density as we increase the deposition pressure during sample preparation. As will be shown below, the resonant soft x-ray data strongly indicate the presence of non-magnetic regions in the high pressure samples.

\section{Magnetometry and Magnetic Force Microscopy}

Prior to the resonant magnetic x-ray scattering, we measured the major hysteresis loops for all of the samples using both magneto-optical Kerr effect (MOKE) and alternating gradient magnetometry (AGM).  The major hysteresis loops, shown in Fig. \ref{fg:hysteresis_loops}, exhibit clear changes with increasing deposition pressure that are related to the increasing structural disorder.  The two lowest pressure samples show classic, low defect thin-film hysteresis curves.  They are characterized by both abrupt domain growth after nucleation and  low remanence.  The higher pressure samples show a constant $dM/dH_{App}$ slope over the majority of the hysteresis loop and reveal increasing coercivity as the sputtering pressure is increased.  The macroscopic nucleation, coercive and saturation fields for each sample are listed in Table 1.  Results from microscopic memory experiments for these samples, as well as additional details of the magnetometry data, were published previously \cite{pierce_prb_07}. The detailed magnetic reversal behavior will be revisited in more detail later in section VI B, when presenting additional high resolution magnetic small angle x-ray scattering data.

\begin{figure}
\includegraphics[width=8.5cm]{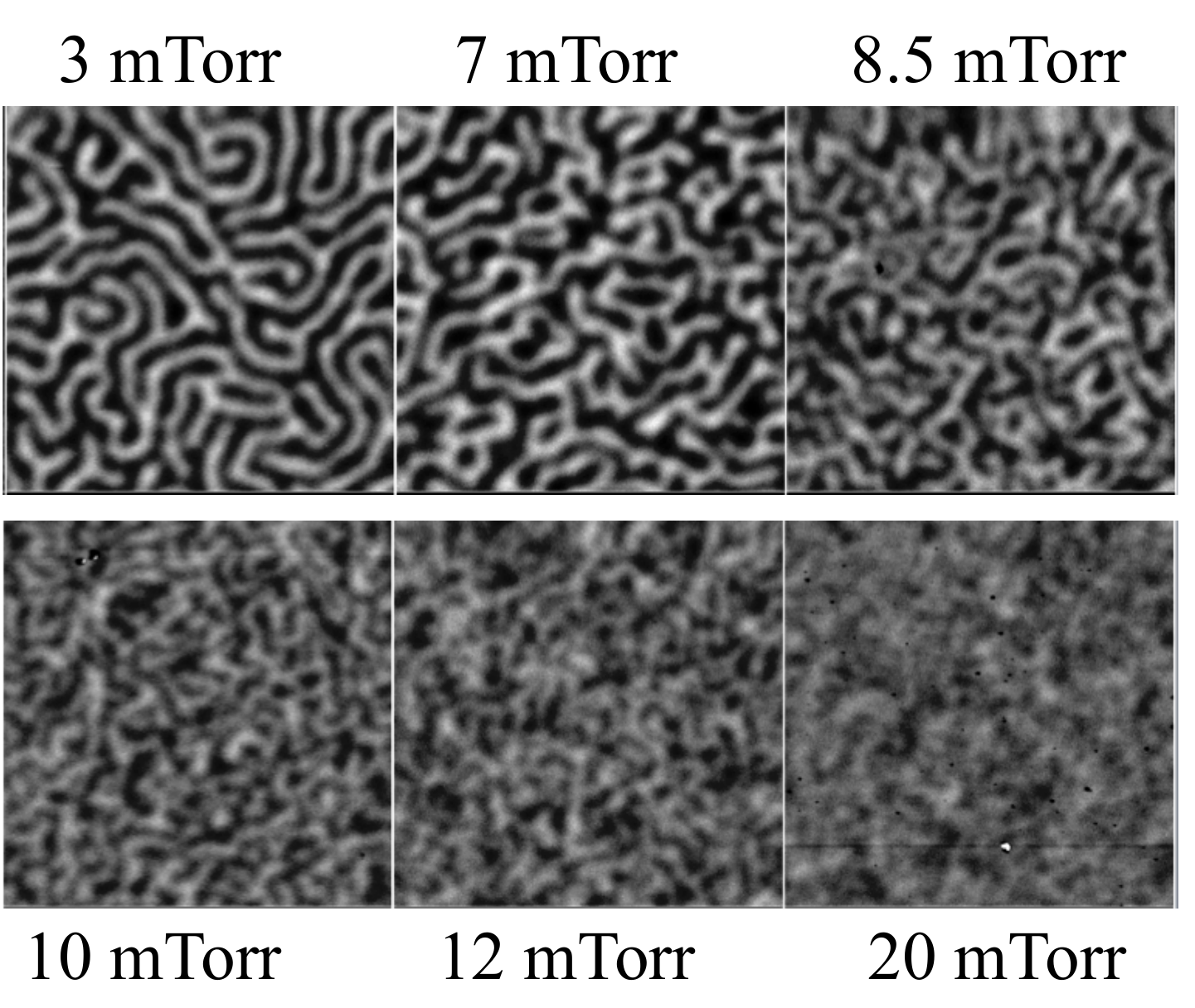}
\caption{(Color Online) MFM images for each sample after out-of-plane AC demagnetization.  Each image is $3\mu m$ on a side.}
\label{fg:mfm}
\end{figure}

We also checked the magnetic domain structure using MFM at remanence after out-of-plane AC demagnetization as shown in Fig. \ref{fg:mfm}.  The domain patterns evolve from well defined labyrinth domains with distinct contrast to a more disordered domain structure with less well-defined contrast and overall smaller domain size. The reduced contrast may result from reaching the resolution limit of the MFM as well as from locally reduced perpendicular anisotropy due to complete Co/Pt layer interdiffusion and loss of Pt (111) texture.  The difficulty in applying large fields during MFM limited our measurements to remanence, however the MFM domain patterns match well with the field dependent resonant x-ray scattering data collected at remanence, as discussed in the following sections.

\begin{table}
\caption{\label{tab:table1}Co/Pt Sample Characteristics }
\begin{ruledtabular}
\begin{tabular}{|l|cc|ccr|}
Sample\footnote{Growth Pressure, mTorr} 
& $\sigma_{RMS}$ \footnote{RMS Surface Roughness from AFM, nm}
& $M_{s}$ \footnote{Saturation Magnetization of Co, $emu/cm^{3}$}
& $H_{n}$ \footnote{Nucleation Field, kOe, from positive saturation}
& $|H_{c}|$ \footnote{Coercive Field, kOe}
& $|H_{s}|$ \footnote{Saturation Field, kOe}\\
\hline
3  & 0.48 & 1360 & 1.58 &  0.16 & 3.7\\
7  & 0.57 & 1392 & 0.64 &  0.68 & 5.0\\
8.5 & 0.62 & 1136 & 1.68 & 1.42 & 5.5\\
10  & 0.69 & 1069 & 1.45 & 1.87 & 6.5\\
12  & 0.90 & 1101 & 1.23 & 2.74 & 9.5\\
20  & 1.44 & 918 & -1.81 & 5.89 & 14.2\\

\end{tabular}
\end{ruledtabular}
\end{table}

\section{Resonant Soft X-ray Scattering from Lateral Charge and Magnetic Heterogeneity}

Several sets of resonant x-ray scattering data near the Co $L_{3}$ and $L_{2}$ edges were collected from this sample series using different detection schemes and beamlines at the Advanced Light Source at Lawrence Berkeley National Laboratory.  One approach used beamlines having high resolution grating monochromators (high longitudinal coherence) and an apertured Si diode detector to measure 1D, extended angular range, radial scans of the azimuthally symmetric transmission scattering patterns as in Ref. \cite{jeff_prb_2001}.  The other used a charge coupled device (CCD) detector and an undulator beam with no monochromator (relaxed longitudinal coherence) to measure the 2D magnetic domain scattering distribution of a more limited angular area directly around the transmitted specular beam. The scattering patterns obtained via this approach exhibit a speckled structure resulting from the use of 35 - 40 micron diameter spatial filters upstream of the samples to increase the transverse coherence of the incident beam, as in Refs. \cite{pierce_prl_03, pierce_prl_05, pierce_prb_07}.   Below the results from these two complementary scattering approaches are presented.  Each approach emphasizes different aspects of the polycrystalline grain structure and the magnetic domain structure, and their variation with pressure and magnetic field.  

For both approaches, data were collected using linear incident polarization that is a coherent superposition of opposite helicity (orthogonal) circular components.  We assume that only the spherically symmetric charge and first order magnetic terms in the scattering factor expansion \cite{hannon_prl_88} contribute to the scattering amplitude.  Theoretical considerations and experimental results \cite{jeff_prb_2001, jeff_prb_05} reveal that intensities measured using linear polarization contain distinct charge-charge and magnetic-magnetic contributions, $I_{cc}$ and $I_{mm}$, respectively, and to a good approximation lack the charge-magnetic cross-term $I_{cm}$.  Interpretations below are based on this understanding.    


\begin{figure}
\includegraphics[width=9.0cm]{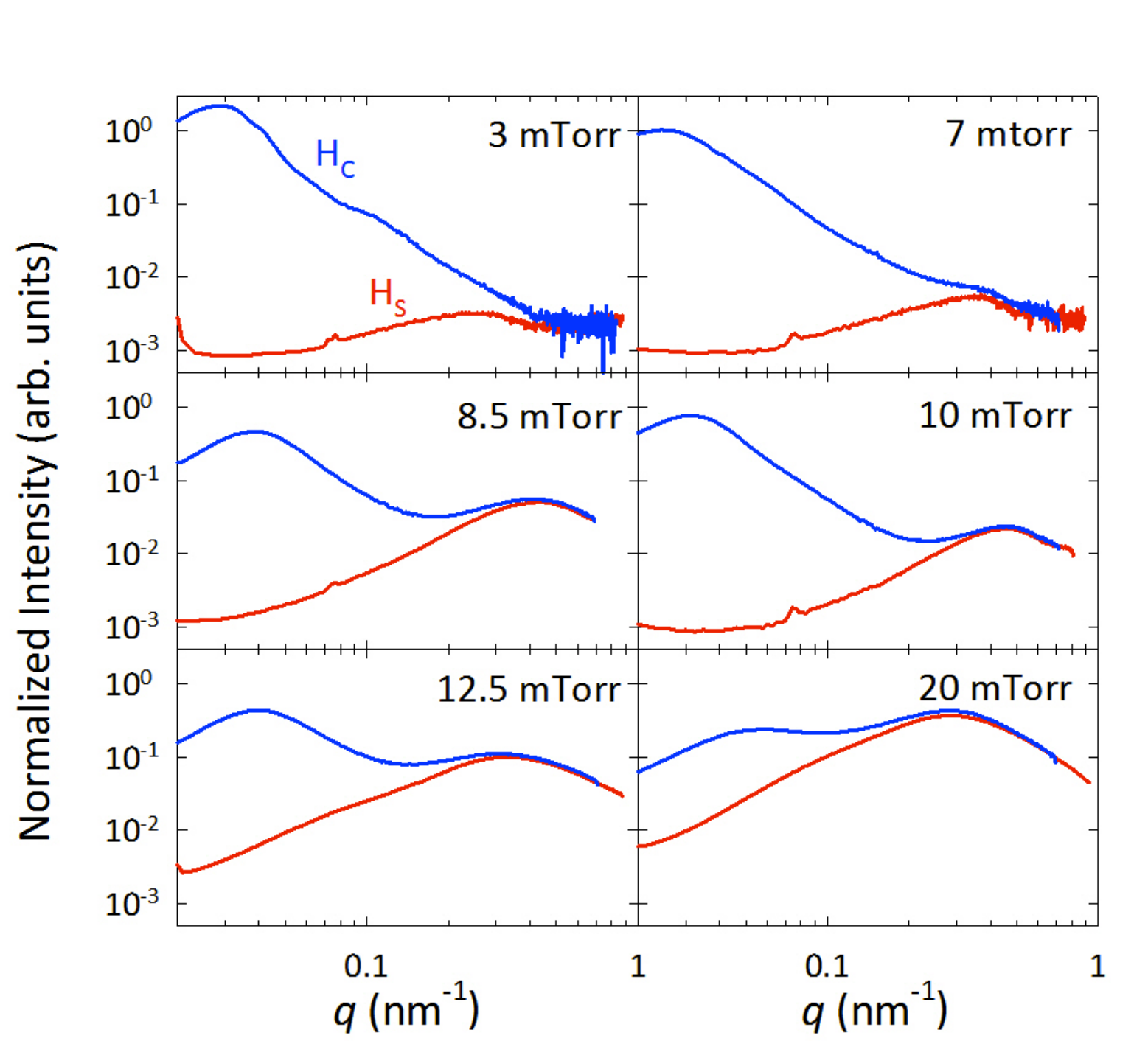}
\caption{(color online) 
Radial transmission scattering profiles measured at the Co $L_{3}$ edge as a function of in-plane scattering vector at saturation field $H_S$ (red) and coercive field $H_C$ (blue). Each sample exhibits a low $q$ peak originating from the magnetic domains and a high $q$ peak originating from structural grains.  Data are normalized to the same intensity scale for comparison.  The weak, sharp feature in the $H_S$ curves at $q \sim$ 0.075 nm$^{-1}$ is a parasitic scattering peak and can be ignored \cite{jk_scattering_note}.  }
\label{fg:sas}
\end{figure}

\subsection{Characterizing magnetic and charge contrast as a function of deposition pressure}

Radial scattering profiles measured at the Co $L_{3}$ edge over an extended q range at saturation field ($H_S$ = 10 kOe) and at the coercive field ($H_C$) of each sample reveal the relative strength and spatial frequency distribution of charge-charge and magnetic-magnetic correlations and how they evolve with growth pressure \cite{jeff_prb_2001}, as shown in Figure \ref{fg:sas}.  All scans were corrected for slit-size to represent the azimuthally integrated intensity vs. radial $q$ and are normalized to the same scale enabling quantitative comparison between samples and scans.  

At $H_C$ magnetic domain scattering is maximized and all samples reveal strong magnetic intensity peaked at low $q$ corresponding to the up-down domain periodicity in the 200 - 400 nm range.  The magnetic domain peak moves to higher $q$, broadens, and weakens as pressure increases.  Finer trends in the magnetic peak during reversal are discussed below.

At saturation magnetic domain scattering is minimized and the remaining intensity results predominantly from pure charge (or chemical) correlations peaking at higher $q$ values corresponding to the polycrystalline grain spatial frequency giving average intergranular spacing ranging from 13 - 24 nm.  The 3 and 7 mTorr samples exhibit weak charge scattering peaks consistent with smoother and more continuous films at this spatial frequency.  As the sample growth pressure increases, the strength of the intergranular peak increases monotonically by 2 orders of magnitude, while its position shifts to higher $q$, reaching a maximum at 10 mTorr before shifting back to lower $q$ as pressure increases further.  

This strong increase in the charge scattering peak with pressure is a direct measure of the decrease in the chemical smoothness and continuity of the films.  This peak, dominated by charge-charge intensity, could have contrast contributions from one or more of the following scattering mechanisms;  surface (or interface) height variation within with the polycrystalline grains, low density or even voided grain boundaries, and chemical segregation at the polycrystalline grain length scale.  Each of these charge contrast mechanisms can reasonably be expected to increase with pressure, consistent with the trends for the $H_S$ profiles in Fig. \ref{fg:sas}.  Further consideration leads to the realization that each of these charge contrast mechanisms implies some associated magnetic scattering contrast.  Thus we expect some $I_{mm}$ contribution at the high $q$ polycrystalline grain peak at saturation in addition to predominant $I_{cc}$ contribution. This is evident in Fig. \ref{fg:sas} as the scattering vector from the coercive field is always larger than the corresponding scattering obtained at saturation, though only by a small amount when shown on a log-scale.  While the low $q$ $I_{mm}$ contribution at $H_C$ results from oppositely oriented domains, at high $q$ and $H_S$ it results from spatial fluctuations in nominally parallel Co magnetization.  

\begin{figure}
\includegraphics[width=9.0cm]{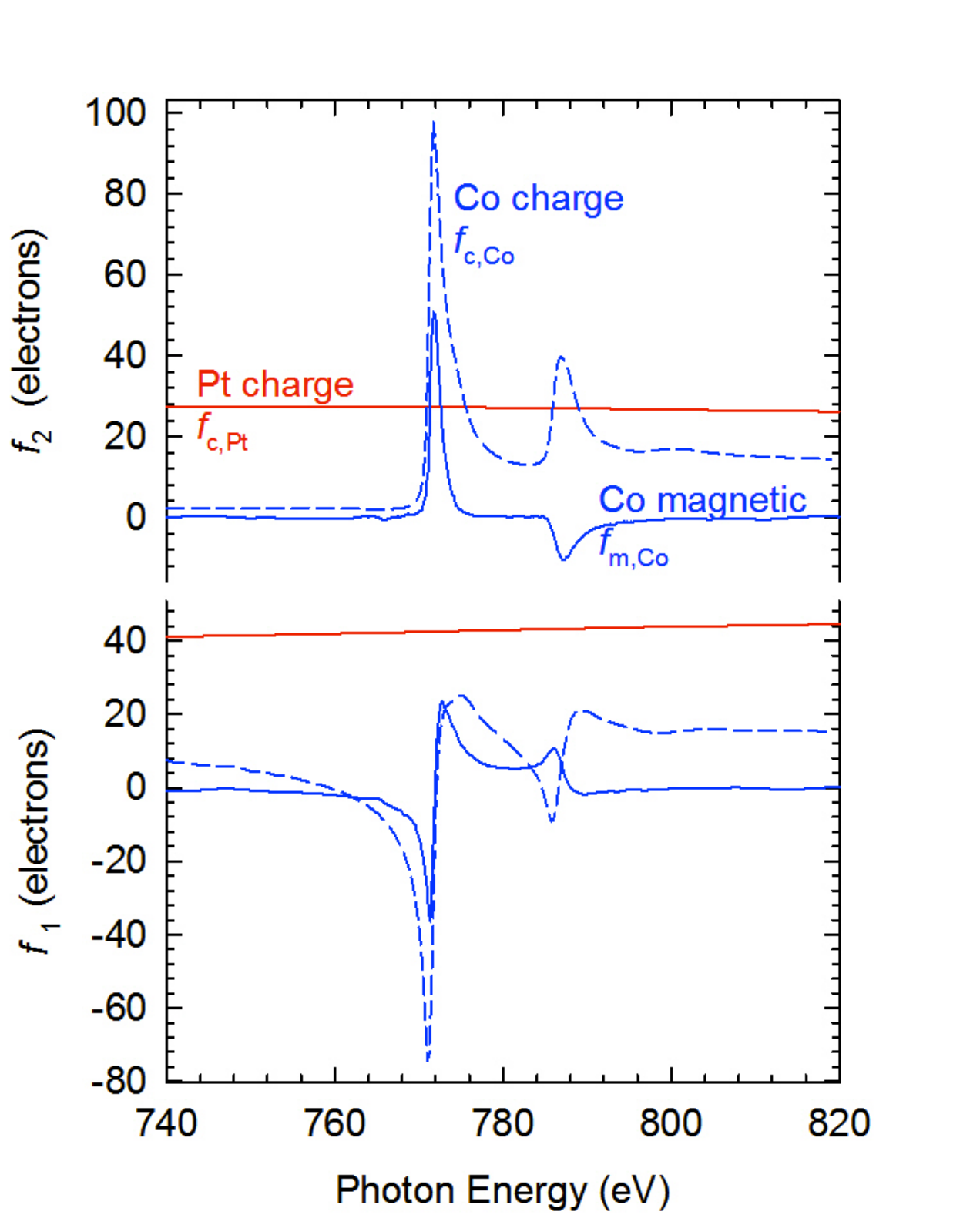}
\caption{(color online) Resonant Co charge (blue, dashed) and magnetic (blue, solid) and non-resonant Pt scattering factors used to model the scattering energy spectra.  $f_1$ represents the real part of the scattering factors and $f_2$ the imaginary part as determined by earlier measurements \cite{jeff_prb_2001}.}
\label{fg:scattering_factors}
\end{figure}

\begin{figure}
\includegraphics[width=9.0cm]{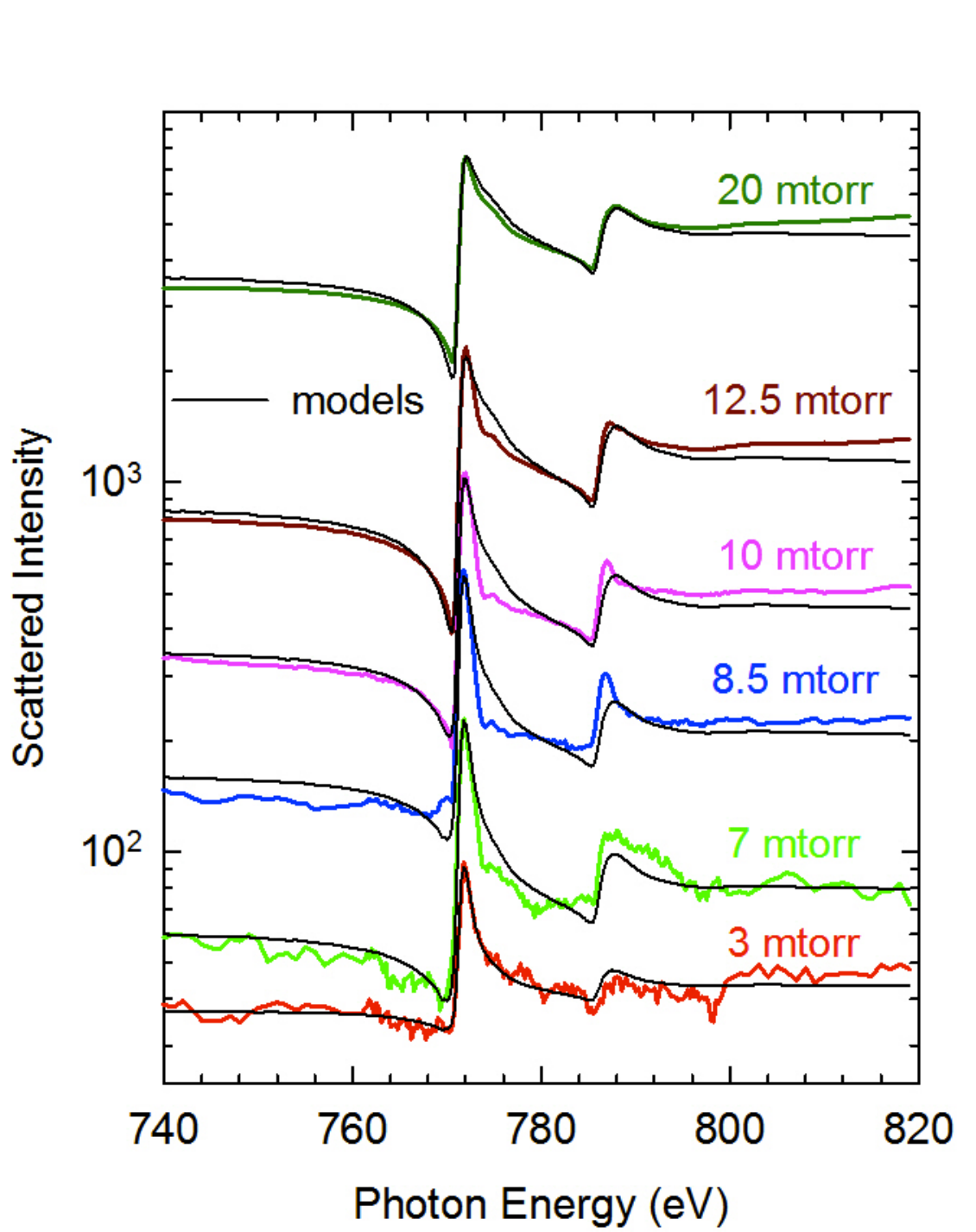}
\caption{(color online) Normalized scattering energy spectra measured using linear polarization at the intergranular peak of each sample.  Black lines through the data are best fits obtained using the model for pure magnetic and pure charge scattering as described in the text.   }
\label{fg:e_spectra}
\end{figure}

With this understanding we measured energy spectra at $H_S$ for the intergrain peaks in Fig. \ref{fg:sas} to quantify the relative contributions of $I_{cc}$ and $I_{mm}$, and their variation with increasing pressure.  The measured spectra are displayed in Figure \ref{fg:e_spectra}, after being normalized to a common scale.  In addition to increasing 2 orders of magnitude in scattering intensity with sample growth pressure, the resonant shape evolves in a systematic fashion as well.  Knowing that resonant Co $I_{cc}$ and $I_{mm}$ spectra have different characteristic shapes, we extend the analysis of Ref. \cite{jeff_prb_2001} to model these energy spectra as a superposition of varying amounts of these two contributions.  For this we use Co resonant charge $f_{Co,c}$ and magnetic $f_{Co,m}$ scattering factors and non-resonant Pt charge scattering $f_{Pt,c}$ as shown in Figure \ref{fg:scattering_factors}.  These data were obtained by averaging Co transmission absorption spectra measured with linear polarization of all samples (they were all nearly identical) yielding the imaginary part of $f_{Co,c}$, and measured x-ray magnetic circular dichroism (XMCD) spectra using circular polarization from one sample yielding $f_{Co,m}$.  The real parts of the scattering factors were obtained via Kramers-Kronig transformation.

\begin{figure}
\includegraphics[width=9cm]{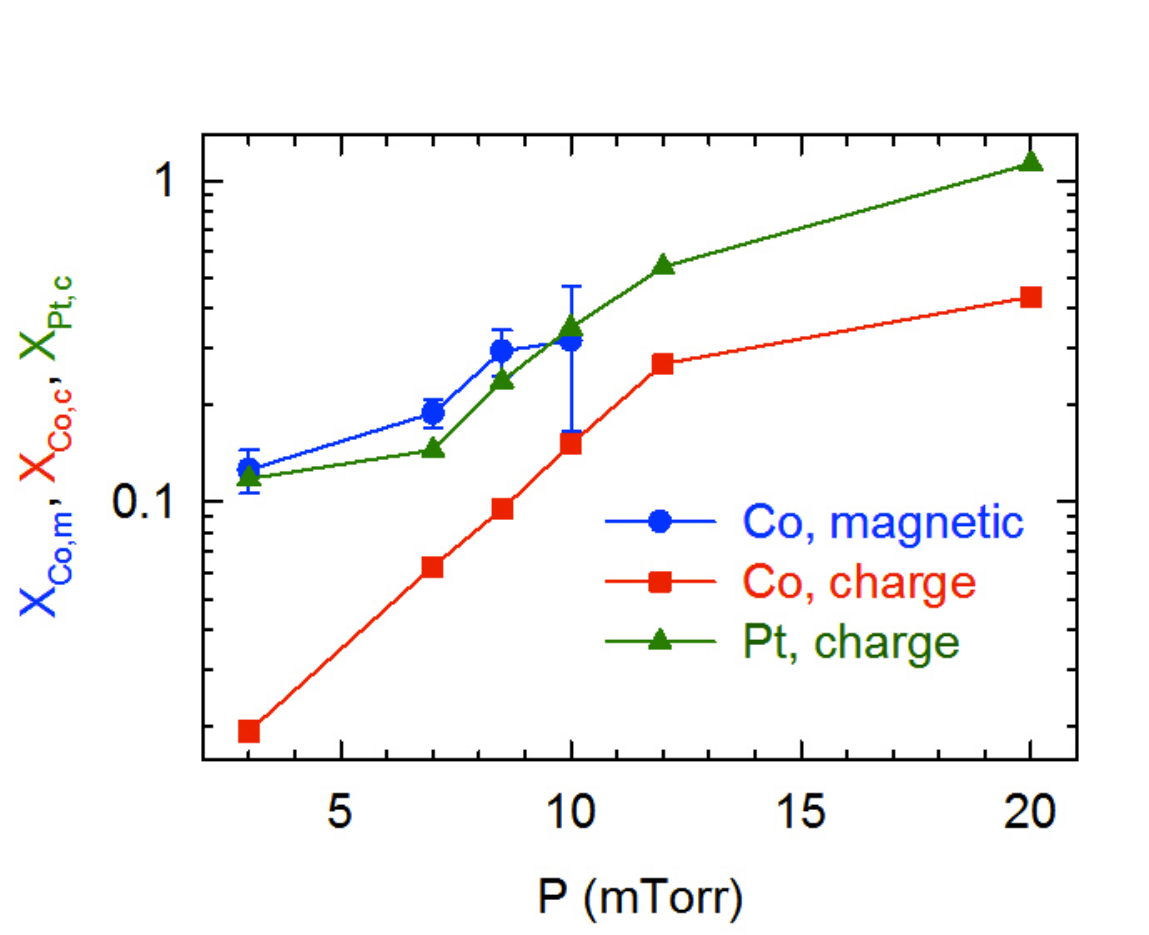}
\caption{(color online) Amplitude weighting factors for Co magnetic (blue circles), Co charge (red squares), and Pt charge (green triangles) scattering factor contributions in the pure magnetic plus pure charge model described in the text. Only the error bars for the Co magnetic contribution are significant and they grow rapidly with pressure, indicating that above 10 mTorr any magnetic contribution is obscured by the more rapidly increasing charge contribution. Lines connect data points. }
\label{fg:e_spectra_result}
\end{figure}

In modeling the scattering spectra as a superposition of $I_{cc}$ and $I_{mm}$, the shape of the charge scattering term must allow for uncertainty in the precise charge scattering mechanism (above) and the evolution of this mechanism with pressure.  Whatever the precise mechanisms we assume only that they can be described as a superposition of $f_{Co,c}$ and $f_{Pt,c}$ scattering factors as these are the only constituents in the sample.  Thus the charge contribution to the intensity is modeled as 
\begin{equation}
I_{cc} = |{X_{Co,c}f_{Co,c} + X_{Pt,c}f_{Pt,c}}|^2
\end{equation}
where the weighting factors $X_{Co,c}$ and $X_{Pt,c}$ give the strength of the Co and Pt charge amplitude contributions.  Since only Co contributes to the magnetic scattering we model this contribution as  
\begin{equation}
I_{mm} = {|X_{Co,m}f_{Co,m} |}^2
\end{equation}
where $X_{Co,m}$ gives the amplitude strength of the magnetic contrast.  We do not constrain $X_{Co,c}$ and $X_{Co,m}$ to be the same, which effectively allows the Co magnetization to vary relative to the Co charge as would be the case if Co in certain regions of the grains has a reduced moment.  The scattering spectra are then described as
\begin{equation}
I = A{(I_{cc} + I_{mm})}
\end{equation} 
where the overall scale factor $A$ is the same for all samples since the data are normalized to a common scale.  

Modeling proceeds by finding the 2 charge and 1 magnetic weighting factors that best fit the measured energy spectra for each pressure.  The best fit models are shown with the data in Figure \ref{fg:e_spectra} and the weighting factors yielding these models are displayed in Figure \ref{fg:e_spectra_result}.  The model spectra generally fit the data well, suggesting that this spectral modeling approach can be used to quantify relative charge and magnetic contributions.  From Ref. \cite{jeff_prb_2001} we expect the $I_{mm}$ and $I_{cc}$ to exhibit characteristic unipolar and bipolar intensity features, respectively, at the Co $L_3$ and $L_2$ lines.  The data and fits both reveal a trend from relatively strong unipolar character at low pressure to predominant bipolar character at high pressure, consistent with relatively stronger $I_{mm}$ contribution at lower pressures for smooth films and predominant $I_{cc}$ contribution at high pressures.  The refined weighting factors confirm this trend, whereby $X_{Co,m}$ is stronger than the charge contributions up until 10 mTorr.  At higher pressures the rapidly increasing charge contributions make it difficult to discern a distinct magnetic contribution, and no magnetic contribution is plotted above 10 mTorr where the error bars become comparable to its value.  

The modeling trends indicate that the charge amplitude increases monotonically by more than an order of magnitude across the pressure range, while the magnetic amplitude starts out relatively strong and increases more slowly before apparently leveling off at 10 mTorr at a value no more than 3 times higher as compared to the 3 mTorr sample.  The continuous and strong increase in charge scattering amplitude with pressure is consistent with the increasing roughness as measured by specular reflectivity and AFM in sections II and III respectively.  While the strong increase in charge scattering amplitude is unmistakable, it is difficult to distinguish between the different mechanisms suggested above from the resonant scattering data alone. 

The slower increase with pressure in the magnetic amplitude followed by apparent saturation can be understood using models in which some portion of the grains, presumably the grain boundaries, become non-magnetic as pressure increases.  This loss of magnetism could result either from large density deficits at the grain boundaries, or from composition variations that would render Co in those regions non-magnetic.  The formation of low density or even void regions at grain boundaries are consistent with earlier presented TEM measurements of the 20 mTorr sample. However, regardless of the specific mechanism, once the magnetism is sufficiently disrupted in some region of the grains, the magnetic scattering would converge towards a constant value, as the magnetic amplitude contrast is then effectively given by $(f_{Co,m} -0)$,  which cannot increase any further.  Thus the trends in $X_{Co,m}$ suggest a loss of magnetism that would tend to magnetically decouple or disrupt adjacent grains, as used for the design of granular magnetic recording media \cite{mcfadyen_mrs_2006, snp_jap_2007}.  Such a disruption in magnetic intergranular exchange with increasing pressure has important implications on the magnetization reversal mechanism and microscopic magnetic memory as measured in the magnetic speckle studies from these samples \cite{pierce_prl_03, pierce_prl_05, pierce_prb_07}.   In particular, magnetically decoupled grains are expected to exhibit more domain memory through hysteresis cycles, as the reversal of individual grains should be more deterministic based on their distribution of anisotropies, Zeeman energies and defects that would act as pinning centers.

\subsection{Resolving fine details of magnetic domain behavior through reversal}

Resonant magnetic scattering at the low q domain peak through magnetization reversal cycles provides further insight into the domain evolution during reversal.  As compared to the previous sections, the results obtained here use a CCD area detector that provides higher angular resolution for a more limited $q$-range around the specular transmitted beam, thus covering only the low $q$ magnetic domain scattering peak earlier observed in Fig. 6.  This domain scattering is dominated by the $I_{mm}$ contribution and the results below ignore the $I_{cc}$ contribution that is small at these low $q$ values.  



The low $q$ resonant magnetic scattering experiments were also performed at the Advanced Light Source at LBNL.  The photon energy was set according to the cobalt $L_{3}$ resonance at
$778$ eV and then adjusted slightly to maximize the difference between scattering at remanence and saturation. The light was passed through a 35 micron diameter pinhole with a sample arm behind the pinhole allowing illumination of a 40 micron diameter area of the sample surface.  
The x-rays are incident perpendicular to the sample surface and are scattered by the sample in a transmission geometry.  The resonant magnetic scattering was then collected by a soft x-ray CCD camera 1.1 meters behind the samples and each scattering pattern was collected with exposures of $\sim$ 10 seconds. The magnetic domains were manipulated by a water cooled electromagnet, which provided {\it in-situ} uniform magnetic fields up to $\sim 11$ kOe  perpendicular to the films.  The experimental chamber was maintained under vacuum at $\sim 10^{-8}$ Torr.

\begin{figure}
\includegraphics[width=7.5cm]{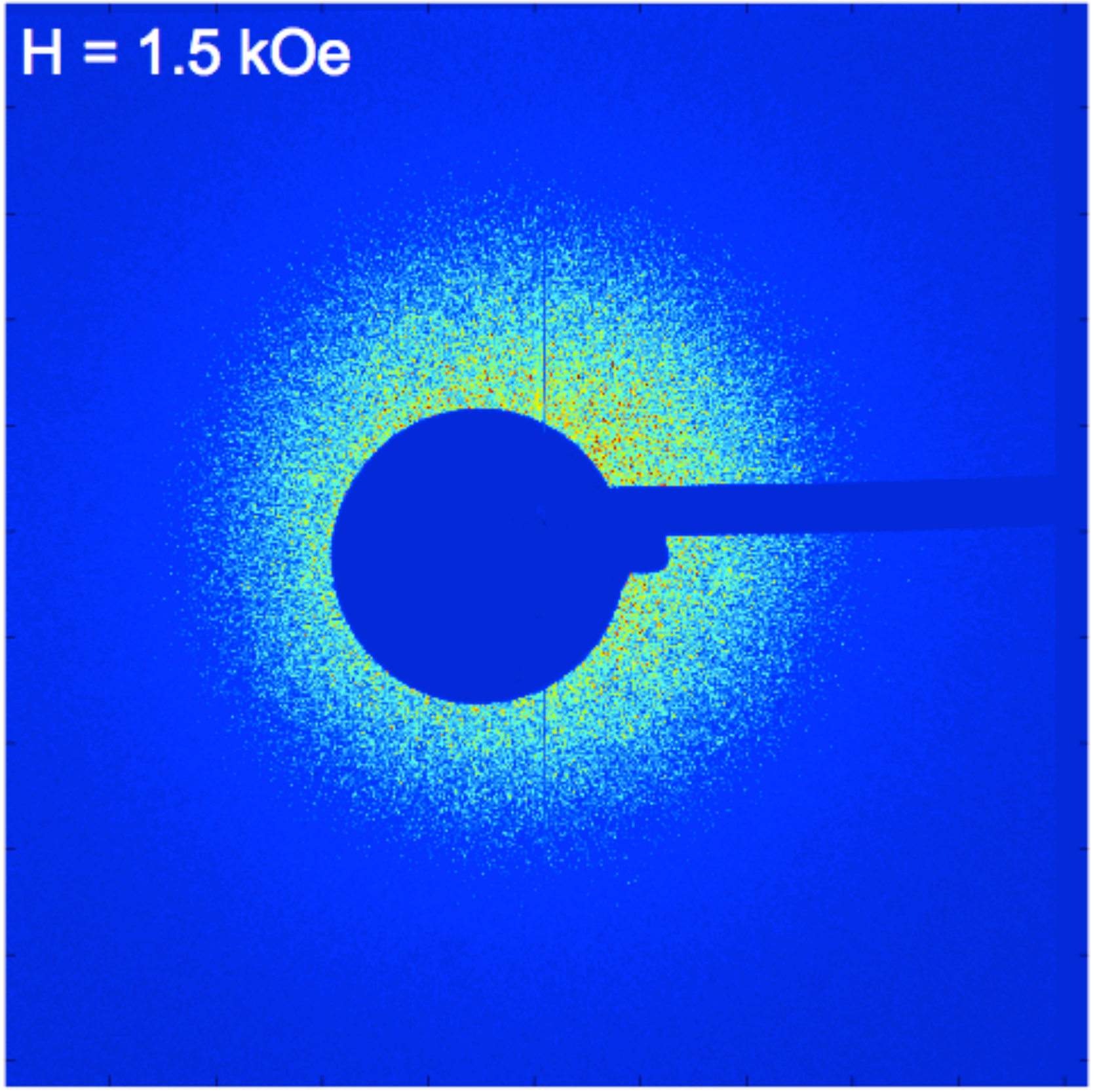}
\includegraphics[width=7.5cm]{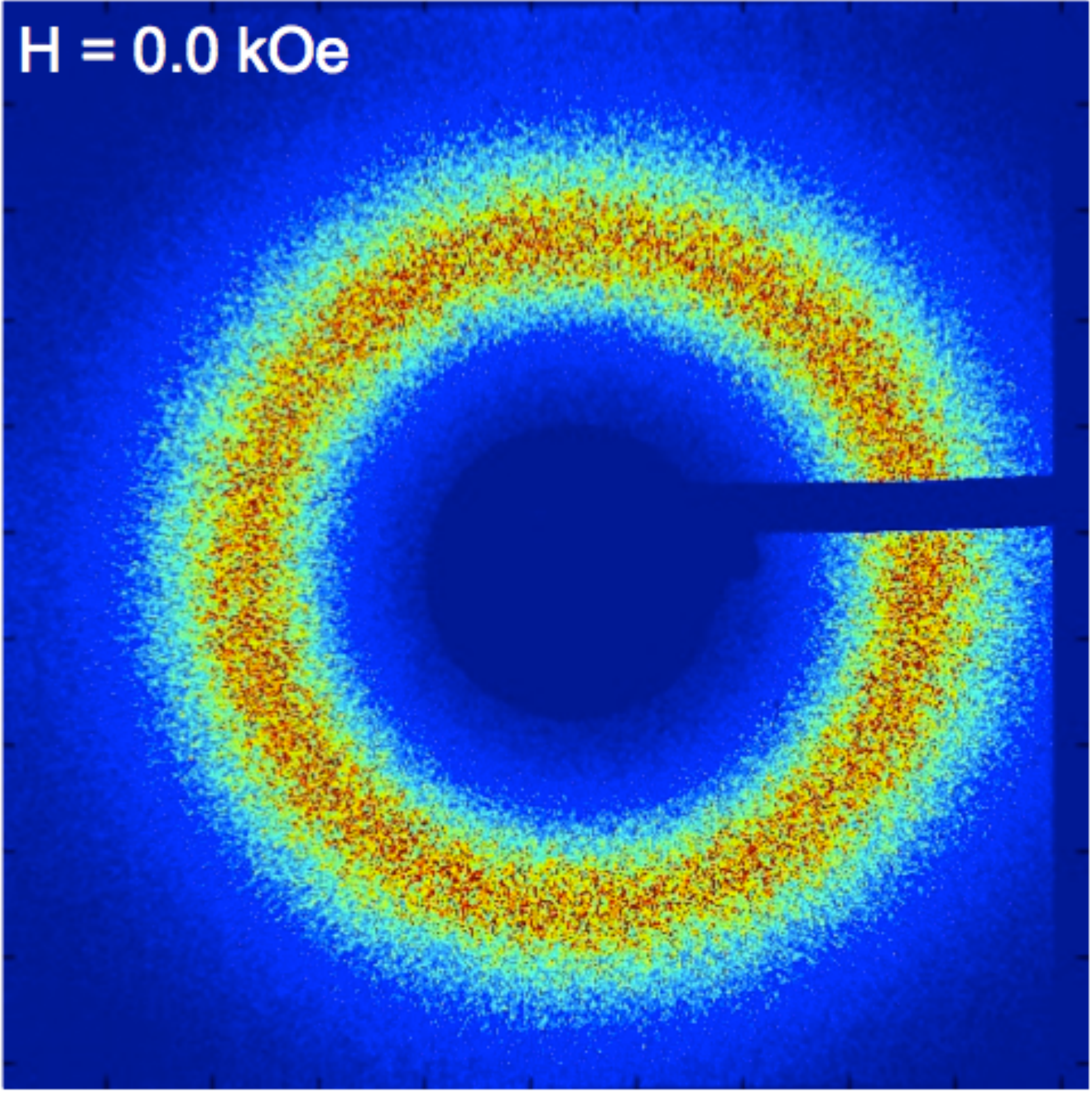}
\caption{(color online) X-ray CCD image for the 3 mTorr sample collected at two different applied fields. The beam-stop is visible in both images.   Top:  Scattering characteristic of a diffuse gas-like phase with randomly distributed magnetic domains.  The scattering intensity rapidly drops with scattering angle.  This image was taken shortly after nucleation for the 3 mTorr sample. Bottom: Image taken at remanence.  The ring of diffuse scattering is characteristic of a liquid-like (labyrinth) arrangement of magnetic domains with first-order positional correlations.  The total observed intensity is also greater at remanence due to the increase in magnetic heterogeneity.
}
\label{fg:ccd_pics}
\end{figure}

Two typical magnetic scattering patterns for the 3 mTorr sample are shown in Figure \ref{fg:ccd_pics}, collected along the major hysteresis loop shortly after nucleation (top) and at remanence (bottom).   At remanence, the dominant structure
is a ring of diffuse scattering reminiscent of the scattering from
a classical 2d liquid exhibiting short-range, nearest neighbor positional correlations, indicating a randomly oriented, labyrinth magnetic domain pattern.  The scattering is azimuthally symmetric about the center of the beam, indicating that there is no preferred alignment direction of the magnetic domains.   The dark feature is the beam-stop and mounting arm.  Note that this diffuse scattering is strongly speckled due to
the use of transversely coherent x-rays.

\begin{figure}
\includegraphics[width=8.8cm]{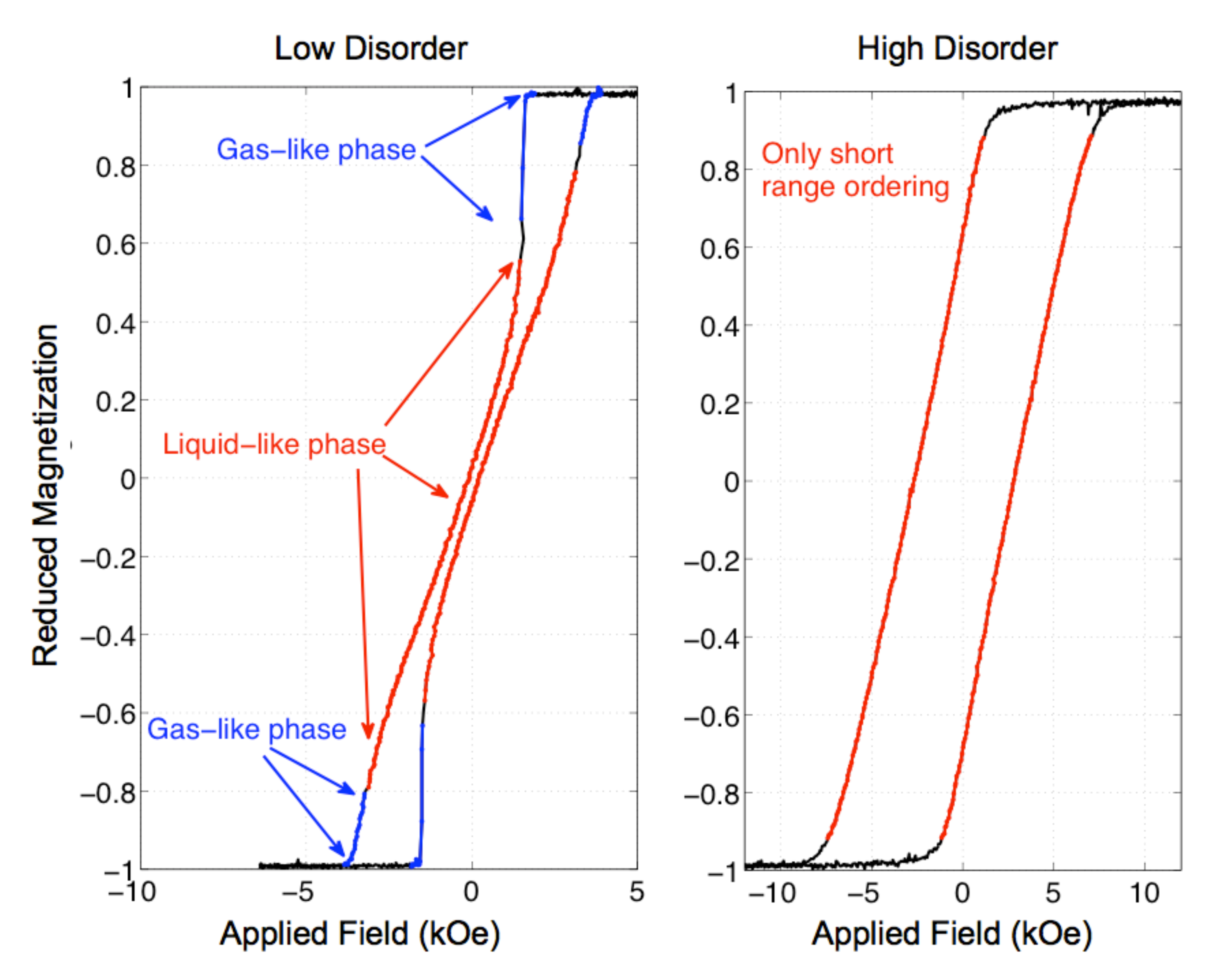}
\caption{(color online) Hysteresis loop plots showing the observed phases for a low pressure (at left, 3mTorr) and  high pressure sample (at right, 12mTorr). In both the 3 mTorr and 7mTorr samples we observed both liquid-like and gas-like scattering.  The gas-like scattering was observed only just after nucleation or just prior to complete reversal. The high pressure samples only showed liquid-like scattering in our experiments.}
\label{fg:phases}
\end{figure}

By changing the applied magnetic field we were able to drive the samples around their major hysteresis loops as well as making excursions inside the major loop.  Each time the applied magnetic field is changed, the domains respond by changing their configuration to minimize the energy of the system.  At each point after a change in the applied field a scattering pattern was collected.  All of these scattering patterns were collected in the quasi-static limit.  After a change in the applied field, the magnetic domains would quickly respond and settle into a new minimum energy configuration. Typically applied field steps of $\Delta H_{App} = 0.25 \pm0.01$ (kOe) were used.

The average width of the magnetic domains is evident in the scattering angle of the diffuse ring.  Likewise the positional correlation of the magnetic domains, the distance over which a domain correlates with its neighbors, is contained in the width of the diffuse scattering ring.  While such information is present in Fig. \ref{fg:sas} for two applied magnetic fields, it was necessary to obtain both higher resolution scattering data and a finer sampling of points along the major hysteresis loop.

Also shown in Figure \ref{fg:ccd_pics} is a speckle pattern collected for the 3mTorr sample shortly after nucleation and the initial growth of the magnetic domains.  In this picture, the scattering falls off exponentially and is reminiscent of the scattering observed from a gas.  The scattering centers are randomly distributed and no definite length scale has developed yet.  Only the two lowest disorder samples exhibited such ``gas-like" scattering patterns with no short range positional order.  For these instances the magnetic domains are randomly distributed over the film and we used a Guinier fit for the exponential decay of the scattering.  This type of scattering pattern was only observed in a narrow region after nucleation or right before saturation on the major hysteresis loop as shown in Fig. \ref{fg:phases}.

\subsubsection{Labyrinth Domains and Liquid-Like Scattering Results
}

The collected scattering patterns contain information about the magnetic domains in the samples.  The angle of maximum intensity, the width of the interference maximum, and the total integrated intensity are all directly related to the magnetic domain structure. 
To extract this information in a systematic fashion ${q} = 0$ was located through a center of mass calculation and the scattering patterns were then averaged azimuthally to eliminate the speckles.  Background images were made by holding the sample at magnetic saturation in a large, constant magnetic field and then subtracted to eliminate small angle charge scattering.

\begin{figure}
\includegraphics[width=9.5cm]{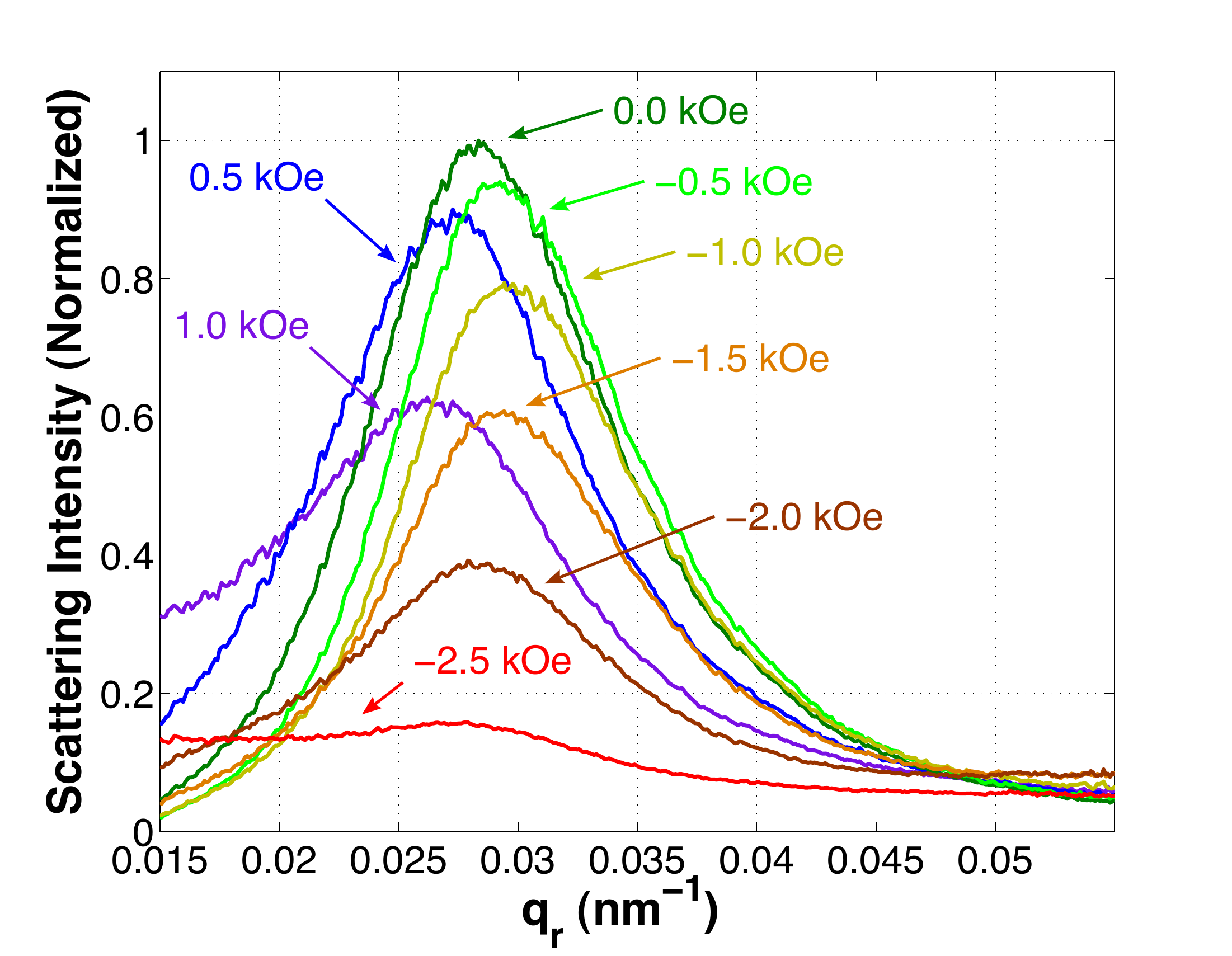}
\caption{(color online) Scattering intensity for the 3 mTorr sample plotted as a function of momentum transfer for several different applied field values in descending order along the major hysteresis loop.  The first applied field shown (1.0 kOe) corresponds to the transition point from gas-like to liquid-like scattering.  The other profiles show further points along the hysteresis loop taken progressively for decreasing applied fields.  The scattering profiles are obtained from azimuthal averages of the SAS CCD images.}
\label{fg:example_profiles}
\end{figure}

A radial profile of the scattering intensity as it depends upon $q_r$ is shown in Fig. \ref{fg:example_profiles} for the 3 mTorr sample.  As shown, the applied field was reduced from a large value taking the samples from positive saturation, down their major loops eventually to negative saturation.   The scattering intensity, peak position, and width all change as the applied field is varied.  Ultimately even the lineshape itself changes as the sample approaches magnetic saturation.  The radius of the scattering annulus from ${q} = 0$ is inversely related to the average domain period via Bragg's law. The width of the scattered ring is inversely related to the range over which neighboring domains are strongly correlated. The greater the variation in width of the domains, the wider the annulus will appear.  As shown earlier in Fig. \ref{fg:sas}, the width of the radial profiles increases with increased interfacial disorder.  And aside from a difference between the 3 mTorr and 7 mTorr samples, the position of the scattering peak shifts towards larger momentum transfer as the disorder is increased.  The larger domain size in the 7 mTorr sample is likely due to the development of initial magnetic point defects that cause some small degree of frustration during an otherwise free stripe domain propagation.

\begin{figure*}
\includegraphics[width=17.5cm]{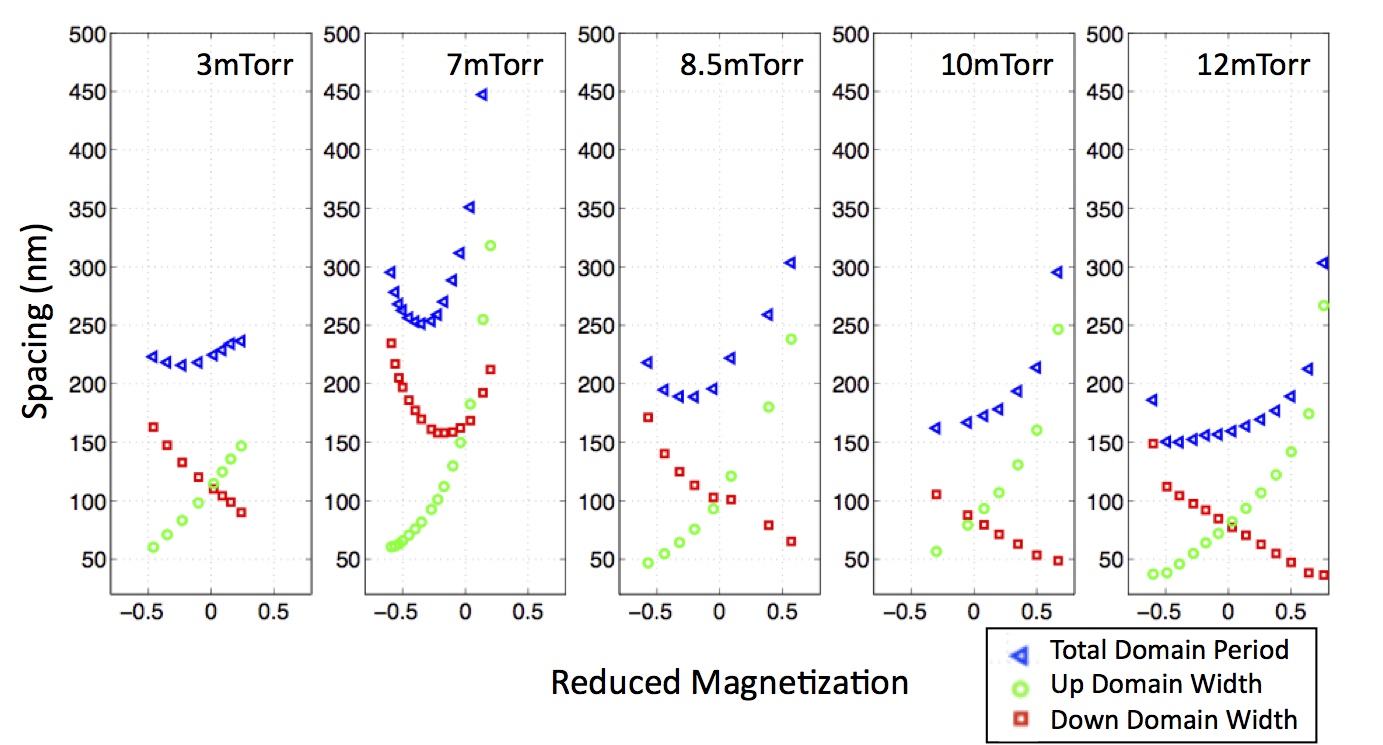}
\caption{(color online) The domain period and the individual up and down domain widths plotted as a function of the reduced magnetization $m = M/M_s$ for the 3, 7, 8.5, 10, and 12 mTorr samples as measured while descending the major hysteresis loop. The widths are based upon a linear decomposition from the sample magnetization.  The domain period $d$ is shown in blue triangles, while $d_{up}$ is shown as green circles and $d_{down}$ as red squares. Values for the 20 mTorr sample are not shown due to the limited $q_r$ range sampled during the high resolution scans. The asymmetry present in each figure will be discussed in the section V.}
\label{fg:all_domain_widths}
\end{figure*}

The radial position of the diffuse scattering is related to the spacing between domains of the same polarity by simple transmission diffraction:  $\lambda = d\sin{\theta}$, with $\lambda = 1.59$ nm, where $d$ is the domain period corresponding to the average distance from a given domain across its anti-aligned neighbor to the next domain of the same polarity. The total domain period dependance on magnetization is shown for the samples in Fig. \ref{fg:all_domain_widths}.    We found that increased sputtering pressure tends to result in smaller domain period.  As disorder increases the domain period shrinks. 

The individual up and down domain widths are of interest as well, but are more difficult to extract from experimental data. Previous x-ray studies \cite{olav_prb_2003} of magnetic domains have fit SAXS intensity data to modeled scattering amplitudes in order to separate the individual up and down domain widths.  Here we propose a much simpler approximation.  So long as the magnetic domains are merely becoming wider and thinner a simple relation between the width of an up domain to the total width is
\begin{equation}
d_{up} = \frac{d_{total}}{2}(1+\frac{M}{M_{s}})
\end{equation}
where $M$ is the current magnetization and $M_{s}$ is the magnetization of the sample at saturation. We will use $m = M/M_s$ for the reduced magnetization below. Figure \ref{fg:all_domain_widths} includes this estimate for the individual domain widths for each sample as they depend upon magnetization going from positive saturation to negative saturation.  This relation will break down once the magnetic domains begin to bifrucate or shrink in length.  At this point the relation will underestimate the shrinking domain width and overestimate the increasing domain width. 

\begin{figure}
\includegraphics[width=9cm]{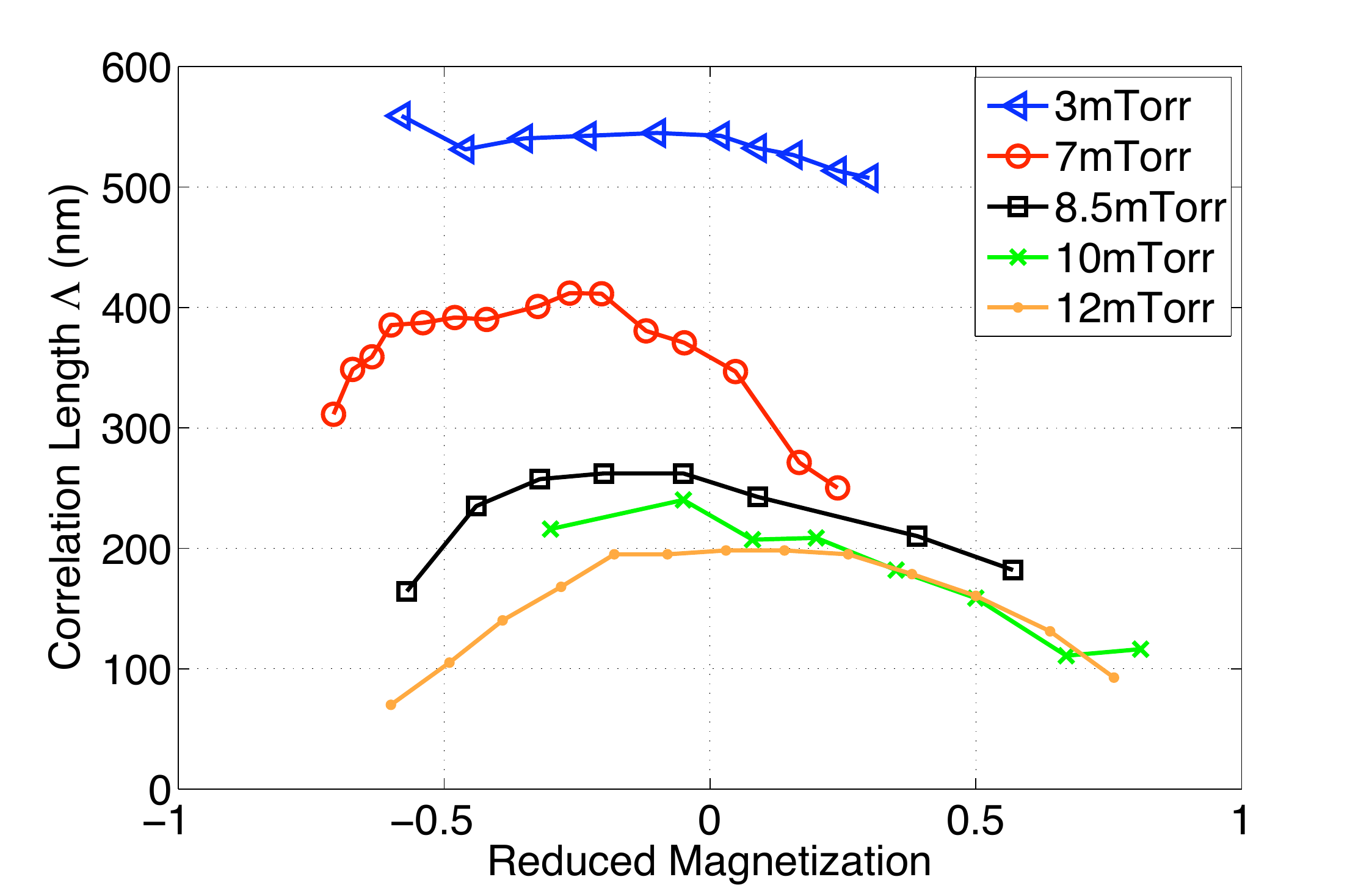}
\includegraphics[width=9cm]{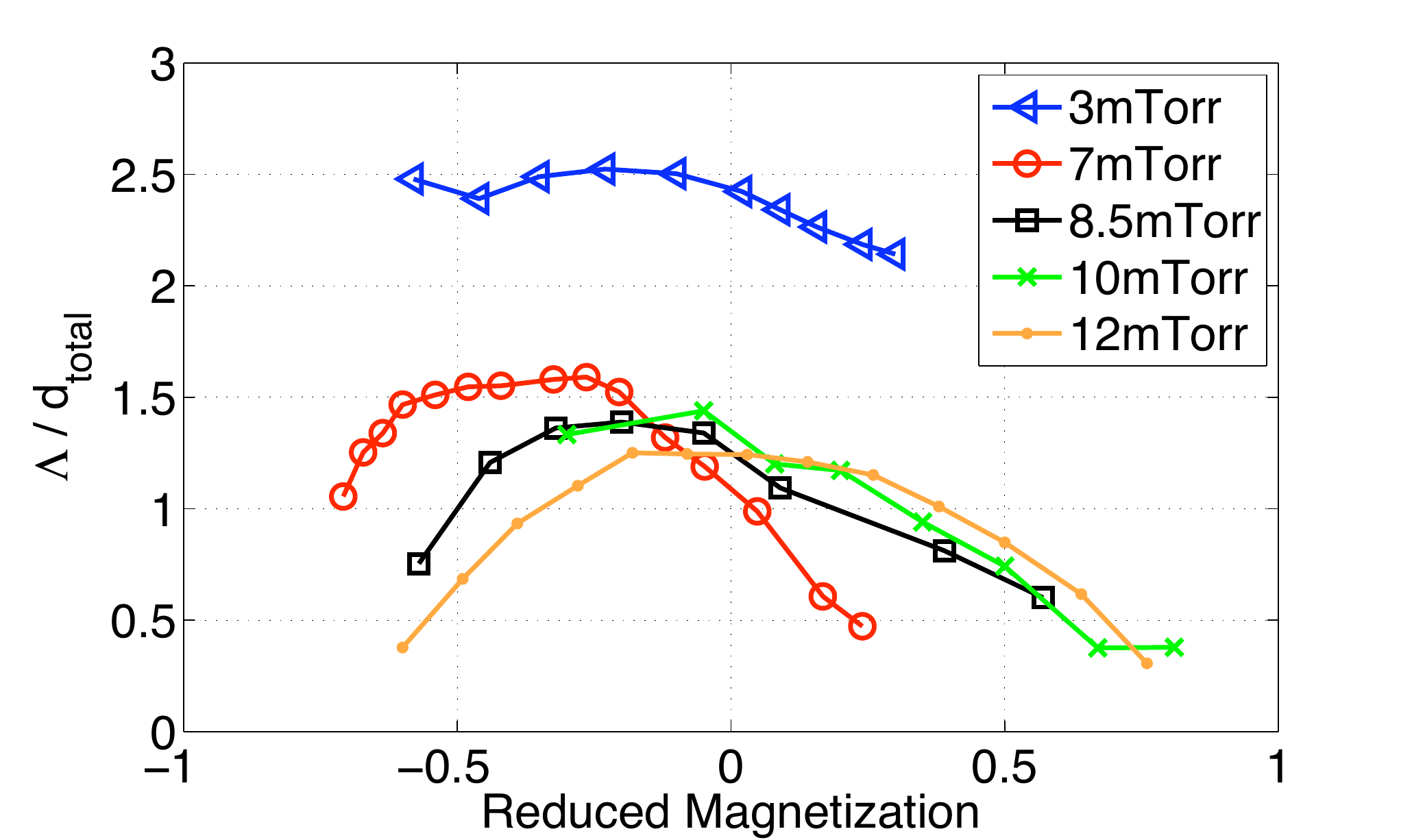}
\caption{(color online) Top: The average correlation lengths for the 3, 7, 8.5, 10 and 12 mTorr samples as they depend upon the applied field around the major hysteresis loop.  The correlation length is related to the inverse of the width of the scattering peak.  Notice that as the disorder is increased the correlation lengths consistently decrease. Bottom : The ratios of the correlation length $\Lambda$ over the total domain period $d_{total}$ plotted as a function of magnetization for each sample.
}
\label{fg:correlation_lengths}
\end{figure}

The degree of correlation between domains, or the relative amount of ordering, for the different samples behaves in a systematic fashion, more so than the domain widths.  Using the real-space 2 point correlation function 
\begin{equation}
C({\mathbf{ r_1} } - {\mathbf{ r_2} }) \sim exp(\frac{-|{\mathbf{ r_1} } - {\mathbf{ r_2} }|^2}{\Lambda^2})
\end{equation}
the correlation length $\Lambda$ can be extracted from the data as well.
Plots of the correlation length as a function of the applied field are shown for the 3, 7, 8.5, 10, \& 12 mTorr samples in Figure~\ref{fg:correlation_lengths} .  For each sample the domain correlation length is initially quite small as the sample nucleates and then grows to a maximum once the domain structure is well established.  Then as saturation is approached the degree of correlation drops again.  Overall the degree of correlation systematically decreases as the sputtering pressured is increased in the samples.   This implies that the random nature of the defects in the samples becomes more and more important compared to the dipole-dipole interaction responsible for the ordering.  

The ratio of the correlation lengths to the total domain spacing $d$, essentially the number of up/down periods within a correlation length, can provide further insight.  The 3 mTorr sample is unique among those studied. The magnitude of the ratios for the 3 mTorr sample is between 2 and 2.5, while for all the other samples the ratio is between 0.5 and 1.5.  The dependence of the ratio upon the magnetization is also quite different.  For the lowest disorder 3 mTorr sample we find that the ratio of correlation length to domain period starts at its maximum value and then decreases as the magnetization is decreased and reversed.  All of the other samples show an initial increase in the ratio of correlation length/domain period with a maximum closer to remanence and a decrease again towards saturation.

\begin{table*}
\caption{\label{tab:table2}Measured Length Scales and Energy Fit Parameters }
\begin{ruledtabular}
\begin{tabular}{|l|c|c|c|c||c|c|c|c|}
Sample\footnote{Growth Pressure, mTorr} 
& Chemical Grain\footnote{nm}
& Magnetic Domain\footnote{nm, at $H_{C}$}
&Correlation Length\footnote{nm, at $H_{C}$}
&Length Ratio\footnote{Correlation Length/Magnetic Domain Spacing}
&$a_{0}$
&$a_{1}$
&$d_{t}$\footnote{nm, from fits}
&$\sigma_{W}/\sigma_{7mTorr}$\footnote{Relative Domain Wall Energy Density}\\
& Period & Period & & & & & & \\
\hline
3 & Not Observed & 216& 542 & 2.5 & 2.23 & .49 & 218$\pm5$ & 0.71\\
7  & Not Observed & 251& 405 & 1.6 & 1.37 & 1.1 & 254$\pm1$  &1\\
8.5  & 16 & 190&259 & 1.4 & 1.71 & .68 & 199$\pm$12 & .38\\
10  &  19& 161 &240 &1.5 & 1.30 & .81 & 165$\pm$10 &.24\\
12  &  26& 150&198& 1.3 & 1.26 & .83 & 145$\pm$8 &  .22\\
20  &  30& 140&144& 1.0& - & - & -& - \\
\end{tabular}
\end{ruledtabular}
\end{table*}

\begin{figure}
\includegraphics[width=8.5cm]{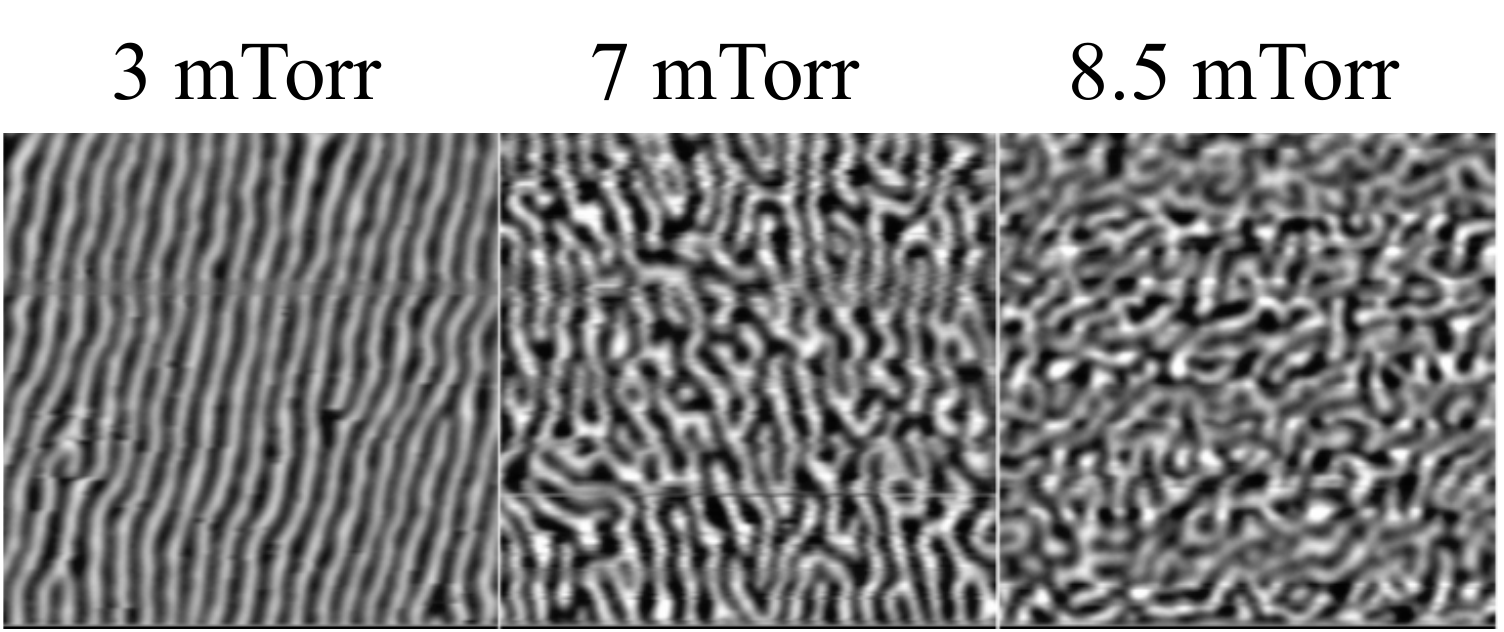}
\caption{MFM images showing the magnetic domain structure resulting from an in-plane demagnetization routine.  Each image is 3 $\mu m$ on a side.}
\label{fg:stripes}
\end{figure}

Additionally, we found with both magnetic x-ray scattering, as well as MFM studies, that the lowest pressure samples could form very regular, long range ordered stripe patterns when in-plane AC demagnetized, i.e. cycled in an alternating positive/negative in-plane field of slowly decreasing amplitude. This behavior shown in Fig. \ref{fg:stripes} was very pronounced in the 3 mTorr sample, already much reduced in the 7 mTorr sample, and barely visible in any of the higher disorder samples.  This possibility of long range parallel domain alignment in the 3 mTorr sample reflects  once more the very low lateral magnetic defect density present within this sample. Even for the 7 mTorr sample it is obvious that a much higher magnetic defect density prevents any long range parallel stripe domain formation after in-plane AC demagnetization, shown in Fig. \ref{fg:stripes}, which is also reflected in the already significantly reduced magnetic correlation length as compared to the 3 mTorr sample.  As expected \cite{kooy}, the domain spacing decreases in the highly ordered parallel stripe patterns.   More extensive work comparing the labyrinth and parallel stripe domain patterns in thin-films has been performed on very similar magnetic samples  \cite{olav_jmmm_07, olav_prb_2003}.

\subsubsection{Gas-like Scattering Results
}

The 3 and 7 mTorr samples both show a very sharp and sudden onset of magnetic reversal thus indicating rapid domain growth initiated by the fact that the Zeemann energy required for nucleating a domain without a magnetic defect present is significantly higher than for propagating an already existing domain. In our scattering experiments, we observed gas like Guinier scattering from the low disorder samples \cite{guinier}.  Both the 3 mTorr and 7 mTorr samples showed evidence of gas-like scattering both shortly after nucleation and just prior to saturation as can be seen in Fig. 16. The corresponding hysteresis loop regions are shown for the 3 mTorr sample in Fig. \ref{fg:phases}.   This is a clear indication that near nucleation and saturation the domains are randomly located within the film with no correlation in spacing or any short range order.  Additionally, upon the appearance of the gas phase after nucleation or just prior to disappearance before saturation, the transition between the liquid-like and gas-like phases was observed, as shown in one of the 3 mTorr datasets included in Fig. 16.  

Simple Guinier fits \cite{guinier} are included in Fig. 16 that describe the data sets well, at times over 2 full decades of intensity.  Using the simplest form where $I(q) \propto e^{-\alpha q^2}$ the data could be fit usually out to $q_r \approx$ 0.03 nm$^{-1}$.  The beamstop prohibited extension to small enough $q_r$ for more detailed modeling.  However, using the most simplistic approach of $I(q) \propto e^{-R_g^2 q^2/3}$ returns radius of gyration values of $R_g$ = 87 $\pm$ 4 nm for the 7 mTorr and $R_g$ = 84 $\pm$ 6 nm for the 3 mTorr samples.  These values were not measurably different for a single sample when comparing the phase just after nucleation to that just prior to saturation, nor do the values show any measurable difference between the 3 and 7 mTorr samples.  Although the SAXS data at these fields did not lend itself to more precise modeling, it is plausible to expect this corresponds to a minimum domain bubble size and minimum domain width of between 60 - 100 nm, depending upon the exact model used, and matches well to bubble domain sizes in similar samples \cite{davies_prb_04, vock}. Likewise it is plausible that for fine enough steps in applied field, that there will be a measurable effect on the $R_g$ value.  However, while the profiles of the scattering are similar, the scattered intensity from the randomly ordered phase of the 7 mTorr sample was several times stronger than the corresponding scattering from the 3 mTorr.  This is an indication that there are more nucleation sites and small isolated domains in the 7 mTorr sample.

Similar X-ray microscopy (XRM) experiments \cite{davies_prb_04} show that the first nucleation sites occur earlier in the hysteresis loop than is often thought. There are isolated bubble domains that appear well before significant domain growth and branching have occurred with sizes that are well below 100 nm. They also confirm that shortly after nucleation the new domains have a fairly constant width, but the distance between nucleated domains is random.  Added to that randomness is the variation for a single nucleated domain that has bent or branched back towards itself.  The distance or spacing between two different parts of that domain is far more random than is observed in a stable labyrinth pattern. Because of combination randomness from the isolated bubble domains, and differing length scales of the isolated domains that have begun to grow, the scattering appears gas-like with no definite length scale.

\begin{figure}
\includegraphics[width=9cm]{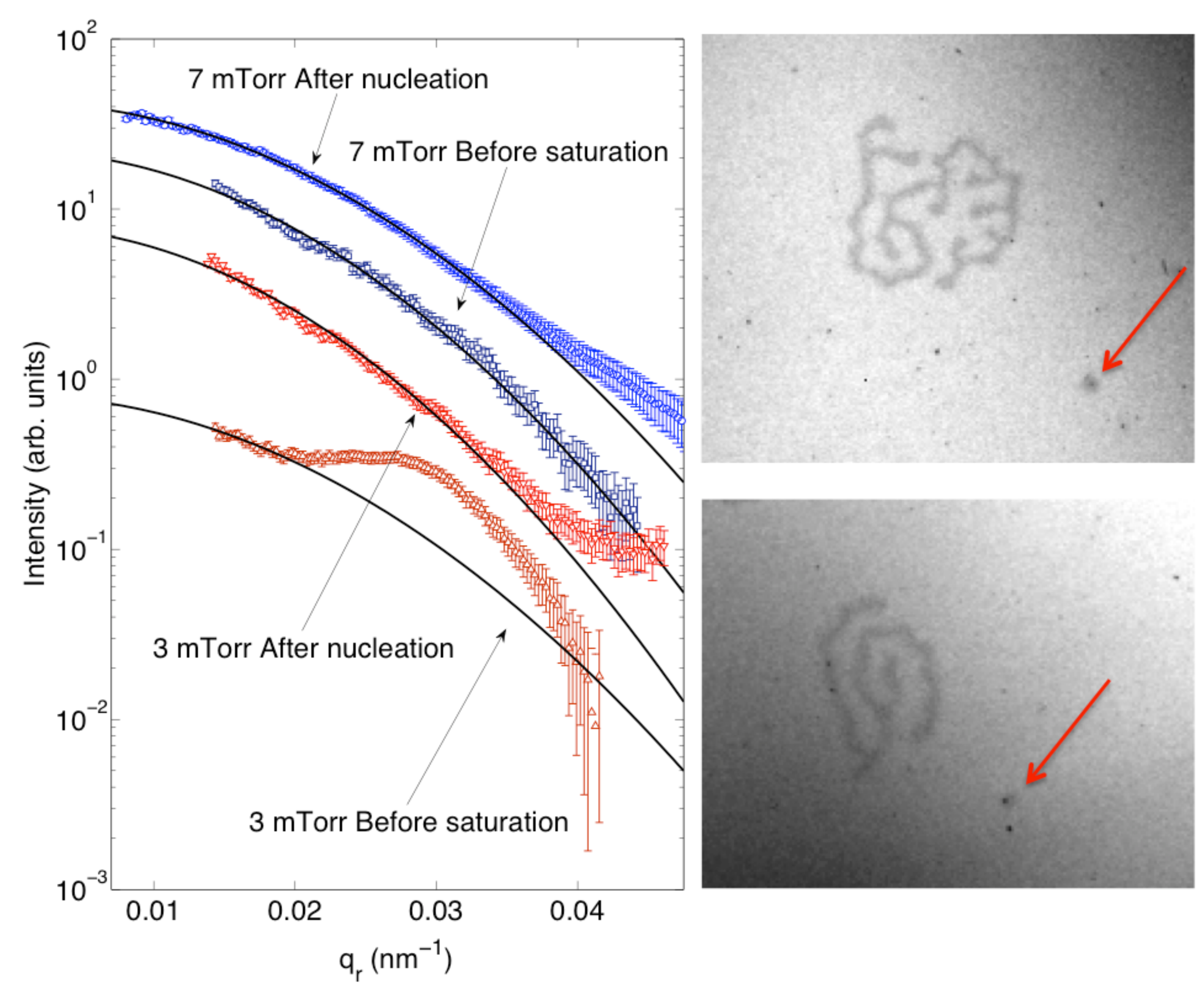}
\caption{(color online) Left : Log-plot of a radial profile for the 7 mTorr (top two) and 3 mTorr (bottom two) samples just after nucleation and just prior to saturation.   The intensity of each set is offset for clarity. Simple Guinier fit lines are included, matching the data at low $q_r$.  The 3 mTorr data just prior to saturation includes the random scattering at low $q_r$, but still shows some evidence of the nearest neighbor positional correlation at higher $q_r$. Right : XRM images of the 3 mTorr sample taken shortly after nucleation. In each image there are nucleated stripe domains which have a definite width, but they are disconnected and not as well defined as in the liquid like state that we observe at remanence.  Isolated bubble domains are also visible with a diameter of $\approx$ 100nm or less \cite{davies_prb_04, vock} as indicated by the arrows.
}
\label{fg:gasphase}
\end{figure}

The random nature of the gas-like phase in the low disorder samples points to fewer nucleation sites and less pinning of the magnetic domain boundaries. Consider a nucleation site which runs out partially, while branching often with some of the branches looping back on themselves. While the width of an individual nucleated domain may be fairly consistent, the width of the region which has not reversed will possess significant randomness until the elongation and branching process has completed. This is further supported by microscopy images as shown in Figure \ref{fg:gasphase} and in other work  \cite{davies_prb_04}.  The absence of such a phase in the higher disorder samples points towards a significant increase in the number of defects that act as nucleation sites.  

This clearly supports the idea of the lowest disorder samples not possessing a chemical grain structure with a distinct grain boundary phase that would act as magnetic defects or pinning sites.   Instead we see the evolution of the magnetic domains to proceed in a fashion consistent with a fluid, i.e. continuous changes in magnetic boundary propagation, where lateral variations in magnetization density or exchange are very small, thus allowing free propagation of magnetic domains into their lowest magnetostatic energy state.

\section{Interpretation of Domain Evolution and Energetics}

The total energy for a continuous magnetic system is often calculated based on the sum of three competing energy contributions: 
\begin{equation}
E = E_{H} + E_{d} + E_{w}
\end{equation}
the energy from the external field $E_{H}$, the energy of the interacting dipoles $E_{d}$, and the energy associated with domain walls $E_{w}$.  Starting from the treatment by Draaisma and de Jonge \cite{draaisma, donzelli} we have modified the equations to include 2 additional phenomenological factors, $a_0$ and $a_1$.  We approximate the dipole-dipole interaction of the domains by
\begin{equation}
E_d = \frac{\mu_0 m^2 M_s^2}{2} + \frac{\mu_0 M_s^2}{2}\sum_{n=1}\frac{4d_{total}}{t(n\pi)^3}\sin^2({\frac{a_1n\pi}{2}(m+a_0)) } 
\label{eq:e_sub_d}
\end{equation}
where $t$ is the total thickness of the magnetic multilayer, $M_s$ is the saturation magnetization, and $m = M/M_s$ the reduced magnetization. With $a_1 = a_0 = 1$ the original equation from ref. \cite{draaisma} is recovered. The factors $a_1$ and $a_0$ allow the domain period $d_{total}(m)$ to be asymmetric with respect to up and down domain portion about $m=0$ and for the minimum $d_{total}$ to occur at non-zero $m$.  We observe both of these features in our data, which are not compatible with the simpler form of the equation for $E_d$.  The parameters $a_0$ and $a_1$ can be qualitatively related to the nucleation field $H_n$, where $a_0$ provides an offset through the phase shift, and slope of the magnetization with applied field, with $a_1$ adjusting the rate of change.   However, a more quantitative relation between them is not possible within our limited data set.
The energy from the external applied field $H$ is written as 
\begin{equation}
E_h = -\mu_0 HM
\end{equation}
and is independent of the quantity of interest, the domain period $d_{total}$.  The cost to create a domain wall is taken to be 
\begin{equation}
E_w = \sigma_w / d_{total}
\label{eq:d_total}
\end{equation}
where $\sigma_w$ represents the energy per unit area of the domain wall.  Therefore, in order to extract the domain period $d_{total}$ the energy $E$ is minimized with respect to $d_{total}$.  The contribution originates from the ratio of the $\partial{E_d}/\partial{d_{total}}$ and $\partial{E_w}/\partial{d_{total}}$ and is given by
\begin{equation}
d_{total} = \sqrt{\sigma_w \mu_0 M_s^2  /  \frac{ \partial{E_d}}{\partial{d_{total}}}}
\end{equation}.

\begin{figure}
\includegraphics[width=8.5cm]{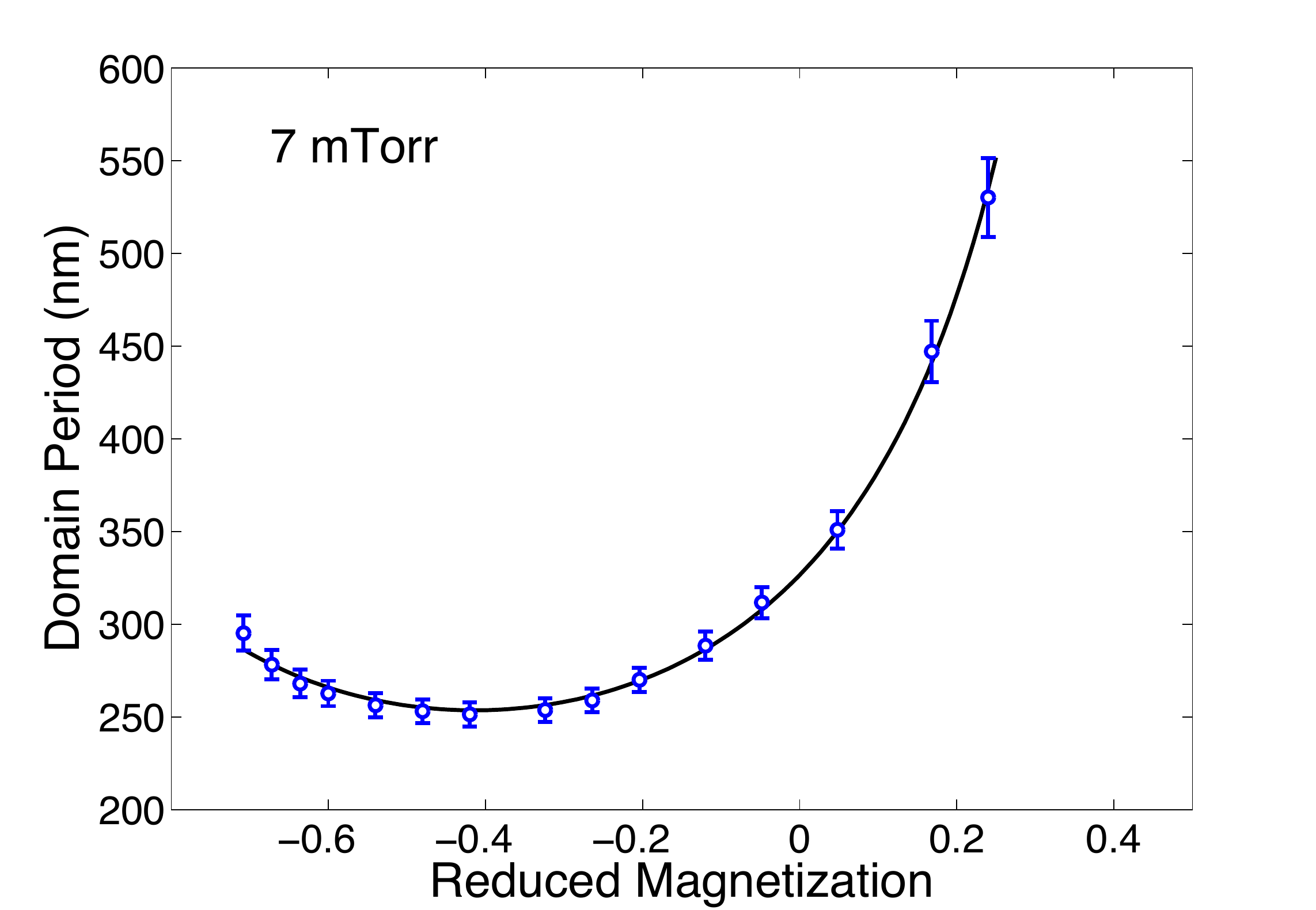}
\caption{(color online) Fit of the domain period as is depends upon reduced magnetization for the 7 mTorr sample.}
\label{fg:ed_fits}
\end{figure}

We carried out the sum for $E_d$ to $n=3$ and fit the resulting equation. An example of the fit to the data is shown in Fig. \ref{fg:ed_fits} for the 7 mTorr sample. The resulting parameters from all the fits, and the calculated domain wall energy density, are listed in Table 2. 
The most interesting of these parameters is $\sigma_w$. For the 7, 8.5, 10, and 12 mTorr we see a steady decrease in $\sigma_w$ as we would expect.  However the energy density of the 3 mTorr sample is between that of the 7 mTorr and 8.5 mTorr samples.   If, as mentioned earlier, we expect the 7 mTorr sample to have an increase in the number of nucleation sites and magnetic point defects, then the rapid domain wall growth after nucleation will lead to a more frustrated domain configuration, and hence a higher domain wall energy associated with it, than for the 3 mTorr sample.  However, once the structural boundaries between chemical grains play an increasingly important role, the energy cost to make a domain would drop since domain walls would form along lateral regions of reduced magnetization.

If we use the fit parameters from equation {\ref{eq:d_total} we are able to reproduce the slopes of the hysteresis curves for the disordered samples, but not the area contained within each curve.  We expect that this is due to the increasing importance of domain wall pinning by grain boundaries of reduced magnetization within the samples.  While the decrease in the domain wall energy density may make it easier for a grain (or a unit area of the film) to flip, it likely impedes domain wall motion itself.  Instead of being able to move freely and continuously in response to an external applied field, the domain walls are tied to the grain boundaries and other defects with reduced magnetization, thus making any domain wall motion less fluid \cite{olav_apl_2011,guenther_apl_08, turner_njp_08}. This is also supported by our earlier magnetic memory experiments performed on the same set of samples  \cite{pierce_prb_07,pierce_prl_05}.  If the domain walls are limited to a more discrete motion along regions of reduced magnetization, such as grain boundaries, then it is also more likely that the microscopic domain evolution could repeat for subsequent cycling along the major hysteresis loop. 

Along similar lines, we can postulate that the correlation length of the domains can be related to the coercivity field $H_c$.  With less disorder (i.e. for lower deposition pressure) we find the domain period to be more uniform and the correlation length to be larger.  As disorder increases due to the formation of well defined grain boundaries and magnetic pinning sites, it will become more difficult for the domain spacing to relax in response to an applied field.  The domain walls will be forced to move along regions of reduced magnetization originating from lateral chemical segregation, thus limiting the distance over which the domains correlate.  How does the domain correlation length compare to the coercivity $H_c$ as the disorder is increased?  We find the inverse of the correlation length to follow $H_c$ very closely as shown in Fig. \ref{fg:hc_vs_cl}.

\begin{figure}
\includegraphics[width=9cm]{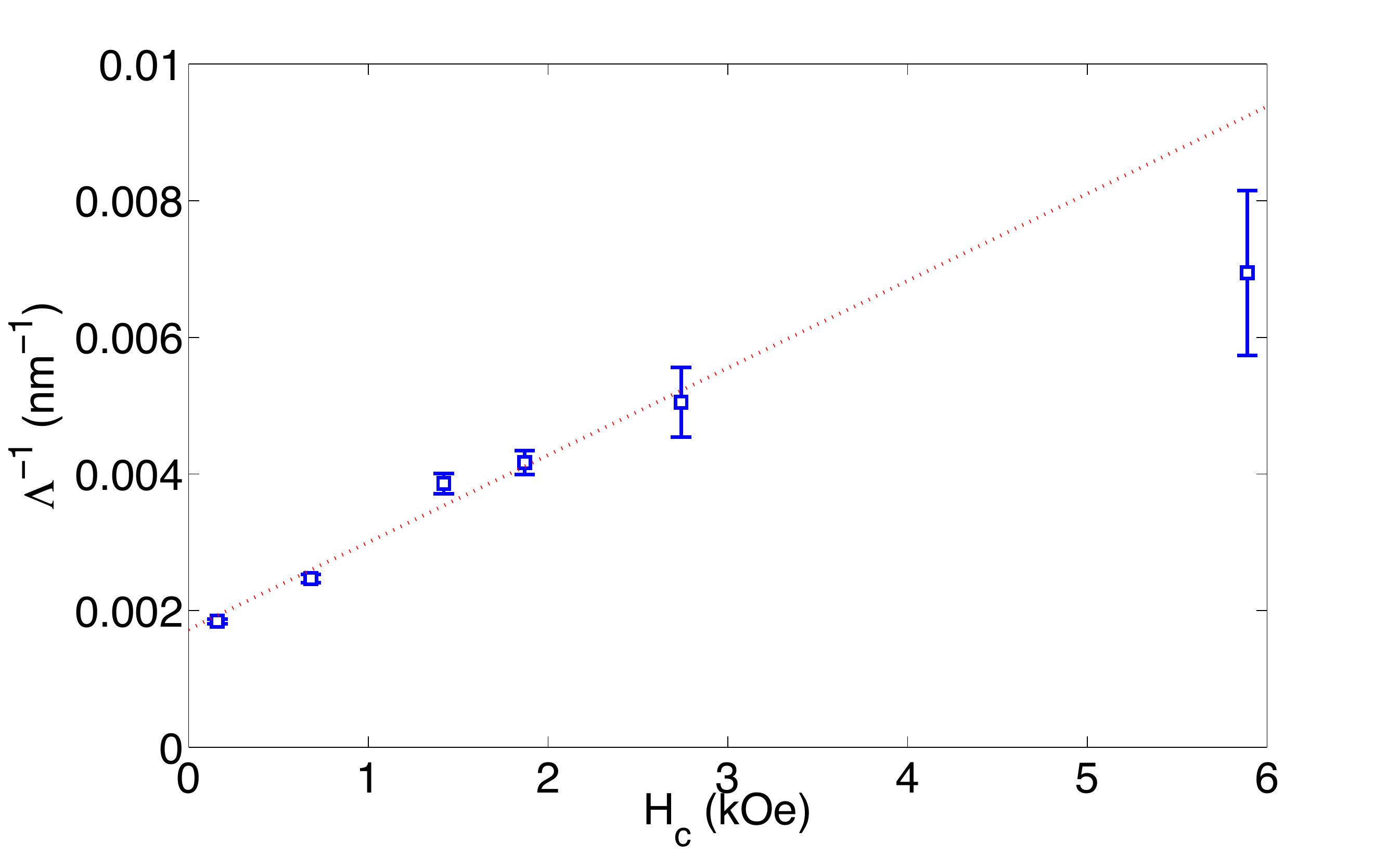}
\caption{(color online) The inverse of the correlation length $\Lambda$ plotted as a function of $H_c$. The fit line shows a simple linear interpolation for the first 5 points. }
\label{fg:hc_vs_cl}
\end{figure}

\section{Discussion}

We can draw several conclusions based upon our observations. The sputtering growth pressure plays a strong role in determining the structure of the magnetic domains and their evolution when applying external magnetic fields.   No evidence of any grain boundary phase is observed in the samples grown at lower pressure.  In contrast, the samples grown at higher pressures exhibit formation of a distinct lower density grain boundary phase in between the grains. This grain boundary phase appears to restrict and discretize the magnetic domain wall motion and overall domain evolution as external fields are applied.  Analyzing the contributions of charge and magnetic scattering of soft x-ray energy spectra taken across the Co L absorption edges at the length scale of the grain period reveal direct evidence that the low density grain boundary phase is associated with an increasingly reduced magnetization density as the deposition pressure is raised from 3 to 10 mTorr.

We have found low structural disorder samples to exhibit hysteresis curves with rapid, avalanche like domain growth, while samples with higher structural disorder reveal a much more gradual onset of magnetization reversal.  The degree of short range ordering of the magnetic domains themselves is also much higher for the low pressure samples as reflected by the domain correlation lengths and in-plane AC demagnetization experiments.  As the number of magnetic domains for a given area increases we know that the gain of creating a new domain wall has gone down. From the individual domain widths it is then possible to estimate the cost of making a domain wall.  From our data we observe a trend towards smaller domain sizes and therefore expect the energy cost of creating a domain wall to be less. This observation can again be explained by the formation of lateral grain boundary phase regions with reduced magnetization density that act as domain wall pinning lines.

With the size of the total domain spacing we see the 7 mTorr to have the largest domain size, followed by the 3 mTorr sample and then all the higher disordered samples with the domain spacing systematically decreasing as the growth pressure increases.  At first this seems not to fit with the simple idea of a systematic dependence of domain width upon disorder.  However it can be understood based upon the number of available nucleation sites.  In the 7 mTorr sample we observe a significant increase in the nucleation site density thus leading to rapid frustration as the domains grow out into larger structures.  For the lowest disorder sample defects that act as nucleation sites are so far apart that any frustration effects are very minor and thus do not influence the domain size as much. In that case, the magnetic properties are only restricted by the competition between magnetostatic and domain wall energy.  Once the number of nucleation sites becomes sufficient to disturb the free magnetic domain evolution, then one consequence is an increased domain size due to frustration. Other studies \cite{davies_apl_09, berger_prb_10} have shown even more dramatic effects that may originate from an increase in nucleation site density. As the disorder is further increased, resulting in the formation of more extended defect lines, the domain period becomes smaller and the correlation length of the domain structures decreases.  By comparing the ratio of the correlation length to the domain period we see that the 3 mTorr sample is distinctly different from all the other samples.

The samples grown at higher pressure, i.e. 8.5 mTorr and above, show evidence of 1-dimensional defect lines that appear in the form of a lower density grain boundary phase.  The magnetic domain walls propagate preferentially along such regions of reduced magnetization, as this allows a significant reduction in domain wall energy. 

All of these properties describe two distinctly different magnetic systems, one disordered and one not.  This result, while being understandable on its own, is further supported by our previous studies of the magnetic memory properties of these samples.  There we found the 3 mTorr and 7 mTorr sample to possess no memory properties, while the 8.5, 10, 12 and 20 mTorr samples all posses significant major loop return point memory.  The 7 mTorr sample shows very little, if any, return point memory.  As such, while the disorder has begun to influence the sample's magnetic properties and microstructure, it is not sufficient for the actual domain configuration to find the microscopically identical reversal path each time the sample is cycled around the major hysteresis loop.

Phenomenological and numerical models \cite{mayergoyz_prl_88, sethna, sethna2, jagla_one, jagla_prb_2005}, in addition to micromagnetics calculations \cite{schrefl_jap_96,weir_jpdap_99},  have been very successful at capturing magnetic domain behavior, as well as memory effects and hysteresis. Other models, such as cellular automata and dynamics simulations, are also capable of reproducing such behavior \cite{mobley_jpcm_04,ren_and_hamley, ren_pre_2001}. We are optimistic that parameter tuning of models which incorporate short and long range interactions, along with a disorder mechanism, will accurately model our observations.

\section{Conclusions}

We have found soft x-ray magnetic scattering to continue to be an excellent technique for characterizing and quantifying the magnetic properties of thin-film samples. Using a model system of perpendicular anisotropy Co/Pt multilayers deposited at different Ar sputter pressure enabled us to continuously tune the degree of lateral heterogeneity and structural disorder in a systematic series of increasingly disordered samples. Combining various advanced characterization techniques, our study allowed new insight into mechanisms of how increased structural disorder and lateral heterogeneity in magnetic thin film systems alters magnetic properties, such as reversal behavior and domain microstructure. 

First we exploit different independent techniques to quantify the degree of structural disorder in our samples, such as surface roughness via AFM, surface and interface roughness via XRR, degree of out-of-plane crystallite alignment via XRD and lateral heterogeneity and grain formation via TEM and resonant soft x-ray small angle charge scattering. The impact of the increasing structural disorder in the form of roughness and distinct grain formation on the magnetic properties of the samples is then studied by hysteresis loop analysis, MFM and resonant soft x-ray small angle magnetic scattering. We find that besides a continuous increase in coercivity with increasing deposition pressure, a crucial transition in magnetic reversal behavior occurs at pressures of about 8-10 mTorr. In this pressure range the mostly free propagation of domains that occurs during the field reversal of low pressure samples is significantly impacted by the structural disorder. As a consequence the typical gas-liquid-gas-like hysteresis loop shape changes into a constant slope reversal mode, where isolated magnetic regions (grains) start reversing more independently from each other, thus also increasing the degree of return point memory in the system as reported in earlier studies \cite{pierce_prl_03, pierce_prl_05, pierce_prb_07}. The initial onset of the transition into a more defect dominated magnetic reversal mode could additionally be revealed via MFM by the possibility of parallel stripe domain alignment after in-plane AC demagnetization, which is already almost completely lost when increasing the deposition pressure from 3 to about 7 mTorr.

Such detailed insight into the impact of structural disorder on magnetic properties and reversal behavior are of great importance for a better understanding of the design of granular magnetic recording media, such as used in hard disk drives, where the controlled segregation of the media layers into a lateral structure of magnetic grains and non-magnetic grain boundaries, has been used for decades to design recording media that sustain ever increasing areal density. While the complex magnetic media layer structure should be maintained within each grain as originally deposited with no inter-diffusion and sharp interfaces between the layers (no disorder here), the desired lateral grain/grain-boundary formation is completely driven by a high deposition pressure self-segregation process of the material, where it is essential to create a laterally heterogeneous structure with a certain degree of disorder. Without such a fine magnetic grain / non-magnetic grain boundary lateral microstructure the media would not be able to confine and preserve (over time) any sharp bit transitions as they are defined by the magnetic write head. We feel that our model system study presented here helps providing a better understanding for the delicate balance between order and disorder that is necessary for designing state of the art magnetic devices based on thin film technology.

We are indebted to numerous people for useful discussions.  We have benefited from numerous helpful discussions with  C. Buechler, J.M. Deutsch, E.A. Jagla,  T. Mai, O. Narayan, and H. You, among many others.   This work was supported by the U.S. Department of Energy, Office of Basic Energy Sciences, Materials Science Division, with particular grants U.S. DOE
via DE-FG02-11ER46831, DE-AC03-76SF00098, and DE-AC02-05CH11231.   Work at UCD was supported by NSF DMR-1008791. OH was partially supported by the Deutsche Forschungsgemeinschaft via HE 3286/1-1.


\begin{references}
\bibitem{by_ref_note}
Note that the 12 mTorr sample had a significantly poorer fit quality than the other pressure samples, which resulted in a slightly increased ML period of 1.13 nm and also may be the reason for the large Co and Pt roughness values observed for the 12 mTorr sample in Fig. 2, which are even larger than for the 20 mTorr sample.
\bibitem{jk_scattering_note}
The parasitic scattering originates from the transmitted portion of the direct beam scattering from the chamber bellows.  Often weak compared to the signals of interest, this scattering is observed even in the absence of a sample.
\end{references}
\end{document}